\documentclass[JEL]{AEA}
\usepackage{natbib}
\usepackage{amsfonts}
\draftSpacing{1.5}
\newtheorem{example}{Example}[section]
\newcommand{\play}{\mathit{play}}
\newcommand{\bbox}{\vrule height7pt width4pt depth1pt}

\newcommand{\commentout}[1]{}

\begin{document}

\title{From outcome-based to language-based preferences}
\shortTitle{}
\author{Valerio Capraro\thanks{Capraro: Department of Economics,
Middlesex University, The Burroughs, London NW4 4BT, U.K.,
v.capraro@mdx.ac.uk.}, Joseph Y. Halpern\thanks{Halpern: Computer Science
Department, Cornell University, Ithaca, NY 14850, USA,
    halpern@cs.cornell.edu.  Work supported in part by NSF grants IIS-178108 and IIS-1703846 and MURI grant
W911NF-19-1-0217.}, Matja\v{z} 
    Perc\thanks{Perc: Faculty of
Natural Sciences and Mathematics, University of Maribor, Koro{\v s}ka
cesta 160, 2000 Maribor, Slovenia \& Department of Medical Research,
China Medical University Hospital, China Medical University, Taichung,
Taiwan \& Complexity Science Hub Vienna, Josefst{\"a}dterstra{\ss}e
39, 1080 Vienna, Austria, matjaz.perc@um.si.}} 
\date{\today}
\pubMonth{}
\pubYear{forthcoming}
\pubVolume{}
\pubIssue{}

\begin{abstract}
    We review the literature on models that try to explain human behavior in
social interactions
described by normal-form games with monetary payoffs.
We start by covering 
social
and moral preferences.
We then focus on the growing body of
research showing that people react to the language in which
actions are described, especially when it activates moral concerns.
We conclude by arguing that behavioral economics is in the midst of a
paradigm shift towards language-based preferences, which will require
an exploration of new models and experimental setups.
\end{abstract}

\maketitle

\section{Introduction}\label{se:intro}

We review the literature on models of human behavior in
social interactions that can be described by normal-form games with
monetary payoffs.
This is certainly a limited set of interactions, since many
interactions are neither one-shot nor simultaneous, nor do they have a
social element, nor do 
they involve just monetary payoffs. Although small, this set of
interactions is of great interest from both the practical and the
theoretical perspective.  For example, it includes games such as
the prisoner's dilemma and the 
dictator game, which capture the essence of some of the most
fundamental aspects of our social life, such as cooperation and
altruism. 
Economists have long recognized that people do not always behave
so as to maximize their monetary payoffs in these games.
Finding a good model that explains behavior has driven much of the
research agenda over the years. 

Earlier work proposed that we have \emph{social preferences}; that is,
our utility function depends not just on our payoffs, but also 
on the
payoffs
of others. For example, we might prefer to minimize inequity or
to maximize 
 the sum of the monetary payoffs,
even at a cost to ourselves.
However, a utility function based on social preferences is still
\emph{outcome-based}; that is, it is still a  function of the players' payoffs. 
We review a growing body of
experimental
literature showing that outcome-based utility functions
cannot adequately describe the whole range of human behavior
in social interactions.
The problem is not with the notion of maximizing expected
utility. Even if we consider decision rules other than maximizing
the expected utility, such as minimizing regret or maximin, we cannot explain
many experimental results as long as
the utility is outcome-based.
Nor does it help to take bounded rationality into account.

The experimental evidence suggests that people have 
\emph{moral preferences}, that is, 
preferences
for doing what they view
as the ``right thing''.  These
preferences cannot be expressed solely in terms of monetary payoffs.
We review the literature, and discuss attempts to construct a utility function
that captures peoples' moral preferences.  
We then consider more broadly the issue of 
language. 
The key takeaway message is
that what 
matters is not just the monetary payoffs associated with actions, but
also how these actions are described. 
A prominent example of this is when the words being used to describe
the strategies activate moral concerns. 


The review is structured as follows. In Section \ref{sec:regularities},
we review the main experimental regularities that were
responsible for the paradigm shift from monetary maximization models
to outcome-based social preferences and then to non-outcome-based
moral preferences. Motivated by these empirical regularities, in the
next sections, we review the approaches that have been taken to
capture human behavior in one-shot
interactions and, for each of them, we describe their strengths and
weaknesses. We start with 
social preferences
(Section \ref{se:social_preferences})
and moral preferences
(Section \ref{se:moral_pref}).
We also discuss 
experimental results showing that the words used
to describe the available actions impact behavior.  In Section
\ref{se:language_based},  we review work on games where the utility
function depends explicitly on the language used to describe the
available actions.
We conclude in Section \ref{se:outlook} with some discussion of
potential  directions for 
future research.   Thinking in terms of moral preferences and, more
generally, language-based preferences suggests potential connections to and
synergies with other disciplines. Work on moral philosophy
and moral psychology could help us understand different types and
dimensions of moral preferences, 
and
work in computational linguistics on 
  sentiment analysis
\citep{pang2002thumbs}
could help explain how we obtain utilities on language.

\section{Experimental regularities}\label{sec:regularities}

The goal of this section is to review a series of experimental
regularities that were observed in normal-form
games played among anonymous players. Although we 
occasionally mention the literature on other types of games, the main
focus of this review is on one-shot, simultaneous-move, anonymous games. 

We start by covering standard experiments in which some people have
been shown not to act so as to maximize their monetary payoff. Then we
move to experiments in which people have been shown not to act
according to any outcome-based preference.
Finally, we describe experiments suggesting that people's 
preferences take into account the words used to describe the
available actions. 

The fact that some people do not act so as to maximize their monetary
payoff was first shown using the 
\emph{dictator game}.
In 
this game,\footnote{Following the standard approach, we 
abuse
terminology slightly and call this a game, although  
it does not specify the utility functions, but only the
outcome associated with each strategy profile. 
}
the \emph{dictator} is given a certain sum of money and has
to decide how much, if any, to give to the \emph{recipient}, who
starts with nothing. The recipient has no choice and receives only the
amount that the dictator decides to give. Since dictators have no monetary
incentives to give, a payoff-maximizing dictator would keep the whole
amount. However, experiments have repeatedly shown that people violate
this prediction
\citep{KKT86,forsythe1994fairness}.
Moreover, the distribution of giving
tends to be bimodal, with peaks at the zero-offer and at the equal share
\citep{engel2011dictator}.

We can summarize the first experimental regularity as follows: 
\begin{quote}
\emph{Experimental Regularity 1.} 
A significant number of
dictators give some money in the dictator game. Moreover, the
distribution of donations tend to be bimodal, with peaks at
zero and at half the total.
\end{quote}

Another classical game in which people often violate the
payoff-maximization assumption is the \emph{ultimatum game}. In its original
formulation, the ultimatum game is 
not a normal-form game, but 
an extensive-form  game, where players move sequentially:
a \emph{proposer} is given a sum of money and has to
decide how much, if any, to offer to the \emph{responder}. The
responder can either accept or reject the offer. If the offer is
accepted, the proposer and the responder are paid according to the
accepted offer; if the offer is rejected, neither the proposer nor the
responder receive any payment. A payoff-maximizing responder clearly
would accept any amount greater than 0; knowing this, a
payoff-maximizing proposer would offer the smallest positive amount
available in the choice set.
Behavioral experiments
showed that people dramatically
violate the payoff-maximizing assumption: responders typically reject
low offers and proposers often offer an equal split
\citep{guth1982experimental,Camerer03}. Rejecting low offers is
impossible to reconcile with a theory of 
payoff maximization.
Making a non-zero offer is consistent with payoff maximization, if a
proposer believes that the responder will reject too low an
offer. However, several researchers
have noticed that offers are typically larger than the amount that
proposers believe would result in acceptance
\citep{henrich2001search,lin2002using}. This led \citet[p. 56]{Camerer03} to
conclude that ``some of [proposer's] generosity in ultimatum games is
altruistic rather than strategic''. These observations have
been replicated in the
normal-form,
 simultaneous-move variant
of the ultimatum game,
that is the focus of this article. In this variant,
the proposer and the responder
simultaneously choose their offer and minimum acceptable offer,
respectively, and then are paid only if the proposer's offer is
greater than or equal to the responder's minimum acceptable
offer.\footnote{Some authors have suggested that the standard
  sequential-move ultimatum game 
  elicits slightly lower rejection rates than its
  simultaneous-move variant
    \citep{schotter1994laboratory,blount1995social}, but this does not
    affect the claim that proposers offer
    more
    than necessary from a purely
    monetary point of view.}

\begin{quote}
\emph{Experimental Regularity 2.} In the ultimatum game, a substantial
proportion of responders reject non-zero offers and a 
significant
number
of proposers offer an equal split.
\end{quote}

The fact that some people do not act so as to maximize their monetary
payoff was also observed in the context of (one-shot, anonymous)
\emph{social dilemmas}. Social dilemmas are situations in which there is a
conflict between the individual interest and the interest of the group
\citep{Hardin68,ostrom1990governing,olson2009logic}. The most-studied
social dilemmas are the \emph{prisoner's dilemma} and the \emph{public-goods
game}.\footnote{Other well-studied social dilemmas are the \emph{Bertrand
    competition} \citep{bertrand1883book} and the \emph{traveler's dilemma}
    \citep{Basu94}. In this review, we focus on
        the prisoner's dilemma and the public-goods game;
the other social dilemmas do not give rise to conceptually
different results, at least within our domain of interest.}

In the
 prisoner's dilemma, two players can either cooperate
(C) or defect (D).
If both players cooperate, they receive
the reward for cooperation,  $R$; if they both defect, they receive the
punishment payoff, $P$; if one player defects and the other
cooperates, the defector receives the temptation payoff, $T$, whereas
the cooperator receives the sucker's payoff, $S$. Payoffs are assumed
to satisfy the inequalities: $T>R>P>S$. These inequalities guarantee
that the only Nash equilibrium is mutual
defection, since defecting strictly dominates cooperating. 
However, mutual
defection gives players a payoff smaller than mutual cooperation.

In the 
public-goods game, each
of $n$ players is given an endowment $e>0$ and has to decide how much,
if any, to contribute to a public pool. Let $c_i\in[0,e]$ be player
$i$'s contribution. Player $i$'s monetary payoff is 
$e-c_i+\alpha\sum_{j=1}^nc_j$, 
where $\alpha\in(\frac1n,1)$ is the marginal return for
cooperation, that is, the proportion of the public good that
is redistributed to each player. Since  $\alpha\in(\frac1n,1)$,
players
maximize their individual monetary payoff by
contributing $0$, but if they do that, they receive less than the amount
they would receive if they all contribute their whole endowment
($e<\alpha n e$). 
%
Although cooperation is not individually optimal in
the prisoner's dilemma or the public-goods game,
many people
cooperate in 
behavioral experiments using these protocols
\citep{rapoport1965prisoner,ledyard1994public}. 

\begin{quote}
\emph{Experimental Regularity 3.} 
A significant number of people
  cooperate in the one-shot prisoner's dilemma and the one-shot public-goods game. 
\end{quote}
  
%
In Section \ref{se:social_preferences}, we show that these
regularities can be explained well by outcome-based preferences, that
is, preferences that are a function of the monetary
payoffs.  But we now discuss a set of empirical findings that cannot be
explained by outcome-based preferences. We
start with truth-telling in tasks in which people can increase their
monetary payoff by misreporting private information. 

Economists have considered several ways of measuring the extent to
which people lie, the most
popular ones being the \emph{sender-receiver game}
\citep{gneezy2005deception} and the \emph{die-under-cup task}
\citep{fischbacher2013lies}. 
In its original version, the
sender-receiver game
is not a normal-form game, but an extensive-form game. There
are
two possible monetary distributions, called Option A and
Option B; only the \emph{sender} is informed about the
payoffs corresponding to these distributions. The sender
can then tell the \emph{receiver} either ``Option A will earn you more
money than 
Option B'' 
or ``Option B will earn you more money than Option
A''. Finally, the receiver decides which of the two options to
implement.\footnote{The original variant of the sender-receiver game
  therefore requires the players to choose their action
    sequentially. 
Similar
results can be obtained with the
normal-form, 
  simultaneous-move variant in which the receiver decides whether to
    believe the sender's message at the same time that the sender
    sends it, or even when the receiver makes no active choice 
    \citep{gneezy2013measuring,biziou2015does}.} 
    Clearly, only one of the
messages that the sender can send is truthful.
\cite{gneezy2005deception} showed that many
senders tell the truth, even when the truthful message does not
maximize the sender's monetary payoff. 
Gneezy also showed that this honest behavior is not driven
by purely monetary preferences over monetary outcomes, since people behaved
differently when asked to choose between the same monetary options when
there was no lying involved
(i.e., when they played a dictator game that was monetarily equivalent
to the sender-receiver game),  
suggesting that people find lying 
intrinsically
costly.

In the die-under-cup task, participants
roll a dice under a cup (i.e., privately) and are asked to report the
outcome. Participants receive a payoff that depends on the reported
outcome, not on the actual outcome. The actual outcome is
typically not known to the experimenter,  but, by comparing the
distribution of the reported outcomes to the uniform distribution,
the experimenter can tell approximately what fraction of
people lied. 
Several
studies 
showed that people tend to be honest, even if this goes against
their monetary interest
\citep{fischbacher2013lies,abeler2019preferences,gerlach2019truth}.

\begin{quote}
    \emph{Experimental Regularity 4.} 
        A significant number of people
        tell the truth in the sender-receiver game and in the
                die-rolling task, even if it lowers 
                their monetary payoff.
\end{quote}

Another line of empirical work that is hard to reconcile with
preferences over monetary payoffs is observed in variants of the
dictator game. For example, \cite{list2007interpretation} explored
people's behavior in two modified dictator games. In the control
condition, the dictator was given \$10 and the receiver was given \$5;
the dictator could then give any amount between \$0 and \$5 to the
recipient. In line with studies using the standard dictator game,
List observed that about 70\% of the
dictators give a non-zero amount, with a peak at \$2.50. In the
experimental condition, List added a ``take'' option: dictators were
allowed to take \$1 dollar from the 
recipient.\footnote{List also considered a
treatment with multiple take options, up to \$5. The results are
  similar to those with a single take option of \$1, so 
  we focus here only on the latter case.} In this case,
List
found that the peak at giving \$2.50
disappears and that the distribution of choices becomes unimodal, with a
peak at giving \$0. However, only 20\% of the participants
choose the take option. This suggests that, for some participants,
giving a positive amount dominates giving \$0 in the baseline, but
giving \$0 dominates giving the same positive amount in the
treatment. This is clearly inconsistent with
outcome-based preferences.   \cite{bardsley2008dictator} 
and \cite{cappelen2013give} 
obtained a
similar result. 

\begin{quote}
\emph{Experimental Regularity 5.} 
A significant number of people prefer giving over keeping in the
standard dictator game without a take option, but prefer keeping over
giving in the dictator game with a take option. 
\end{quote}

In a similar vein, \cite{lazear2012sorting} showed that some
dictators give part of their endowment when they are constrained to
play a dictator game, but, given the option of 
receiving the maximum amount of money they could get by playing the
dictator game without actually playing it, they choose to
avoid the interaction.  Indeed, \cite{dana2006you} found that  
some dictators would even pay \$1 in order to avoid the interaction.
Clearly, these results are inconsistent with
outcome-based preferences.

\begin{quote}
\emph{Experimental Regularity 6.} 
A significant number of people prefer giving over keeping in the
standard dictator game without an exit option, but prefer keeping over
giving in the dictator game with an exit option. 
\end{quote}

How a game is framed is also well known to affect people's behavior.
For example, 
contributions in the public-goods game depend on whether the game
is presented in terms of positive externalities or negative ones
\citep{andreoni1995cooperation}, rates of cooperation in the 
prisoner's dilemma depend on whether the game is called ``the community
game'' or ``the Wall Street game'' \citep{ross1996naive}, and 
using terms such as ``partner'' or ``opponent'' can affect
participants' behavior in the trust game
\citep{burnham2000friend}. \cite{burnham2000friend} suggested that the
key issue (at least, in these 
situations%
) is what players perceive as the
norms.

Following this suggestion, there was work exploring
the effect of changing the 
norms
on people's
behavior. One line explored dictators' behavior in variants of the
dictator game in which the initial endowment is not given to the
dictator, but is instead given to the recipient, or equally shared between
the dictator and the recipient, and the dictator can take some of the
recipient's endowment; this is called the ``take'' frame. 
Some experiments showed that people tend to be more altruistic in the
dictator game in the ``take'' frame than in the standard dictator game
\citep{swope2008social,krupka2013identifying}; moreover,
this effect is driven by the perception of what the socially
appropriate action is \citep{krupka2013identifying}.
However, this result was not replicated in other experiments
\citep{dreber2013people,eckel1996altruism,halvorsen2015dictators,hauge2016keeping}.
A related stream of papers pointed out that including morally loaded
words in the instructions of the dictator game can impact dictators'
giving
\citep{branas2007promoting,capraro2019power,capraro2019increasing,chang2019rhetoric},
and that the behavioral change can be explained by a change in the
perception of what dictators think the morally right action is
\citep{capraro2019power}.
Although there is debate
about whether the ``take'' frame can
impact people's behavior in the dictator game, 
there is general agreement that
the wording of the instructions can impact dictators' behavior by
activating moral concerns. 

The fact that the wording of the instructions can
impact behavior has also been observed in other games.
For example, \cite{eriksson2017costly} found
that the language 
in which the rejection option is presented significantly impacts
rejection rates in the ultimatum game. Specifically, receivers are
more likely to decline an 
offer when this option is labelled ``rejecting the proposer's
offer'' than when it is labelled ``reducing the proposer's
payoff''. Moreover, in line with the discussion above regarding the
dictator game, Eriksson et al.~found this effect to be
driven by the perception of what the morally right action is.

A similar result was obtained in the \emph{trade-off game}
\citep{capraro2018right,tappin2018doing}, where a
decision-maker has to unilaterally decide between two allocations of
money that affect the decision-maker and two other
participants. One allocation equalizes the payoffs of the three
participants; the other allocation
maximizes the sum of the payoffs, but is
unequal. \cite{capraro2018right} and \cite{tappin2018doing} found that
minor changes in the language in which the actions are presented
significantly impacts decision-makers' behavior. For example, naming
the efficient choice ``more generous'' and the equitable choice ``less
generous'' makes subjects more likely to 
choose the efficient choice, while naming the efficient choice ``less
fair'' and the equitable choice ``more fair'' makes subjects more likely
to choose the equitable choice.

Morally loaded words have also been shown to affect behavior in the prisoner's
dilemma, where participants have been observed to cooperate at different rates 
depending on whether
the strategies are named `I cooperate/I cheat' or `A/B' \citep{mieth2021moral}.
Furthermore,
in both the one-shot and iterated prisoner's dilemma,
moral suasion%
, that is, providing participants with cues that make the morality of an action salient,
increases
cooperation
\citep{capraro2019increasing,dal2014right}.
This suggests that cooperative behavior is partly driven by a
desire to do what is morally right. 

\begin{quote}
\emph{Experimental Regularity 7.} Behavior in several experimental
games, including the dictator game, the prisoner's dilemma, the
ultimatum game, and the trade-off game, depends on the instructions
used to introduce the games, especially when they activate moral
concerns. 
\end{quote}

\section{Social preferences}\label{se:social_preferences}

In order to explain the seven regularities listed in Section
\ref{sec:regularities}, one has to go beyond preferences for
maximizing monetary payoffs.
The first generation of utility functions that do this appeared in the 1970s.
These \emph{social preferences} share the underlying assumption that
the utility of an individual depends  not only on the individual's monetary
payoff, but also on the
monetary payoff of the other players involved in the
interaction.\footnote{There has also been work on 
explaining human behavior
in terms of
bounded rationality. The idea behind this approach is that 
computing a best response may be computationally
difficult, so
players do so only to the best of their ability. 
Although useful in many domains, these models do not
explain deviations from the payoff-maximizing strategy in situations in
which computing this strategy is obvious, such as
in the
dictator game (Experimental Regularity 1). While some people
may give in the dictator game because 
they incorrectly computed the payoff-maximizing
strategy, did not read the instructions, or played
randomly, this does not begin to
explain the overall behavior of dictators. In a recent
analysis of over 3,500 dictators, all of whom had correctly answered a
comprehension question regarding which strategy maximizes their
monetary payoff, \cite{branas2018gender} found an average donation of
30.8\%. Similar observations also apply to the other experimental
regularities listed in Section \ref{sec:regularities}.}
 In
this sense, these social preferences are all particular instances of
\emph{outcome-based preferences}, that is, 
utility functions that depend only on (1) the
individuals involved in the interaction and (2) the monetary payoffs
associated with each strategy profile.
Social preferences
typically explain Experimental Regularities 1--3 well.
However, they have difficulties with Experimental Regularities 4--7.
In this section, we review
this line of work. 
For more comprehensive reviews, we refer the readers to
\cite{Camerer03} and \cite{dhami2016foundations}. 
Part of this ground was also covered by \cite{sobel2005interdependent}.

Economists have long recognized the need to include
other-regarding preferences in the utility function. Earlier work
focused on economies with one private good and one public good. In
these economies, there are $n$ players; player $i$ is endowed with 
wealth $w_i$, which she can allocate to the private good or the
public good. This is a quite general class of 
games 
(e.g., the
dictator game, 
prisoner's dilemma, and 
the 
public-goods game can
all be expressed in this form), although it does not cover several
other 
games 
of interest for this review (e.g., the 
ultimatum
 and trade-off games). 
Let $x_i$ and $g_i$ be $i$'s allocation to
the private good and contribution to the public good,
respectively. Economists first assumed that $i$'s
utility
$u_i$
depended only
and monotonically
on $x_i$ and $G=\sum
g_j$. 
According to this model, the government forcing an increase of contributions to public goods (e.g., by increasing taxes) will result in a decrease of private contributions, dollar-for-dollar.  Specifically, if the government takes one dollar from a particular contributor and puts it in the public good,
while keeping everything else fixed (say, by changing the tax
structure appropriately), then that contributor can restore
the equilibrium by reducing his 
contribution
by
one dollar 
\citep{warr1982pareto,roberts1984positive,bernheim1986voluntary,andreoni1988privately}%
. 
The prediction that this would happen
was violated in empirical studies
\citep{abrams1978crowding,abrams1984crowding}. Motivated by these
limitations, \cite{andreoni1990impure}
introduced a theory of \emph{warm-glow giving}, where the utility
function captures the intuition that individuals receive 
positive utility  from the very act of giving to the public
good. Formally, this corresponds to considering, instead of a utility
function of the form $u_i=u_i(x_i,G)$, one of the form
$u_i=u_i(x_i,G,g_i)$. 
Note that the latter utility function is still outcome-based, because
all of its arguments 
are functions of monetary outcomes.
Warm-glow theory has been applied successfully to
several domains. However, when it comes to explaining the experimental
regularities listed in Section \ref{sec:regularities}, it has 
two significant limitations. The first is its domain of applicability:
the 
ultimatum 
and trade-off games
cannot be expressed
in terms of economies with one private good and one public good in any
obvious way. The second is more fundamental: as we show at the
end of this section, it cannot explain Experimental
Regularities 4-7 because it is outcome-based.

More recently, economists have started defining the utility function
directly on the monetary payoffs of the players involved in the
interaction. These utility functions, by construction, can be applied to any
economic interaction. The simplest such utility
function is just a linear combination of the individual's payoff and
the payoffs of the other players
\citep{ledyard1994public}. Formally, let $(x_1,\ldots,x_n)$ be a
monetary allocation among $n$ players. The utility of player $i$
given this allocation is 
$$
u_i(x_1,\ldots,x_n)=x_i+\alpha_i\sum_{j\neq i}x_j,
$$
where $\alpha_i$ is an individual parameter representing $i$'s level
of altruism.
Preferring to maximize payoff is the special case where 
$\alpha_i=0$.
Players with $\alpha_i>0$ care positively about the payoff of
other players. Consequently, this utility function is consistent with 
altruistic behavior 
in the dictator game and with 
cooperative behavior in social dilemmas. Players 
with $\alpha_i<0$ are spiteful. These are players who prefer to maximize
the differences between their own monetary payoff and the monetary
payoff of other players. Thus, this type of utility function is also consistent
with the existence of people who reject positive offers in the
ultimatum game. 

However, it was soon observed that this type of utility function is
not consistent 
with the quantitative details of ultimatum-game experiments. Indeed,
from the rate of rejections observed in an experiment, one can easily
compute the distribution of the spitefulness parameter. Since
proposers are drawn from the same population, one can then use this
distribution to compute what offer should be made by proposers.
The offers that the proposers should make, according to the $\alpha$
calculated, are substantially larger than those observed 
in the experiment \citep{levine1998modeling}.

Starting from this
observation, \citet{levine1998modeling} proposed a generalization of 
Ledyard's utility function that assumed that 
agents 
have information (or beliefs)
about the level of altruism of the other players and base
their own level of altruism on that of the other players.
This allows us to formalize the intuition that players may 
be more altruistic towards altruistic players than towards
selfish or spiteful players. Specifically, Levine
proposed the utility function 
$$
u_i(x_1,\ldots,x_n)=x_i+\sum_{j\neq i}\frac{\alpha_i+\lambda\alpha_j}{1+\lambda}x_j,
$$
where $\lambda\in[0,1]$ is a parameter representing how sensitive
players are to the level of altruism of the other players. If
$\lambda=0$, then we obtain Ledyard's 
model, where $i$'s level of altruism towards $j$
does not depend on $j$'s level of altruism towards $i$; if
$\lambda>0$, players tend to be more altruistic towards altruistic
players than towards selfish and spiteful
players. 
Levine
showed that this model fits 
the empirical data in several settings quite well, including the 
ultimatum
game and prisoner's dilemma.

One year later, \citet{fehr1999theory} introduced a utility function based on a
somewhat different idea. Instead of caring directly about the absolute
payoffs of the other players,
Fehr and Schmidt
assumed that (some) people care
about 
minimizing payoff differences. Following this intuition, they
introduced a family of utility function that can capture inequity aversion: 
\begin{equation*}\begin{split}
u_i(x_1,\ldots,x_n)&=x_i\\
&-\frac{\alpha_i}{n-1}\sum_{j\neq i}\max(x_j-x_i,0)\\
&-\frac{\beta_i}{n-1}\sum_{j\neq i}\max(x_i-x_j,0).
\end{split}\end{equation*}
Note that the first sum is greater than zero if and only if
some player $j$ receives more than player $i$
($x_j>x_i$). Therefore, $\alpha_i$ can be interpreted as a parameter
representing the extent to which player $i$ is averse to having player
$j$'s payoff higher than his own.
Similarly, 
$\beta_i$ can be interpreted as
a parameter representing the extent to which player $i$ dislikes
advantageous inequities.
Fehr and Schmidt
also assumed that
$\beta_i\leq\alpha_i$ and $\beta_i\in[0,1)$. The first assumption
means that players dislike 
  having a payoff higher than that of some other player at most as much
  as they dislike having a payoff lower than that of 
  some other player.
  To understand the assumption that 
$\beta_i < 1$,
suppose that player
  $i$ has a 
    payoff larger than the payoff of all the other players. 
Then $i$'s utility function reduces
  to
$$
u_i(x_1,\ldots,x_n)=(1-\beta_i)x_i+\frac{\beta_i}{n-1}\sum_{j\neq i} x_j.
$$
If $\beta_i\geq1$, then the component of the utility function
corresponding to the monetary payoff of player $i$ is non-positive,
so player $i$ maximizes utility by giving away all his money,
an assumption that seems implausible \citep{fehr1999theory}.
Finally, the assumption $\beta_i\geq0$ simply means that there are no players
who prefer to be better off than other
players.\footnote{\cite{fehr1999theory} make this assumption for
simplicity, although they acknowledge that they believe that there
  are subjects with $\beta_i<0$.} 
This
way of capturing inequity aversion has been shown to fit empirical
data well in a 
number of contexts, including the standard ultimatum game, variants
with multiple proposers
and with multiple responders,
and the 
public-goods game.

A different type of utility function capturing inequity aversion was
introduced by 
\citet{bolton2000erc}. Like Fehr and Schmidt
(\citeyear{fehr1999theory}), Bolton and Ockenfels assume that players'
utility function 
takes into account inequities among players. To define this function,
they assume that the monetary payoffs of all players
are non-negative. Then they define
$$
\sigma_i=\begin{cases}
\frac{x_i}{c} \qquad\text{if $c>0$,} \\
                        \frac1n \qquad\text{ if $c=0$,}
                    \end{cases}
$$
where $c=\sum_{j=1}^nx_j$, so $\sigma_i$ represents $i$'s
relative share of the total monetary payoff. 
Bolton and Ockenfels further assume that $i$'s utility function (which they
refer to as $i$'s \emph{motivational function}) depends only on 
$i$'s monetary payoff $x_i$ and on his relative share $\sigma_i$, 
and satisfies four assumptions. We refer to
\citet{bolton2000erc} for the formal details; here we focus on an
intuitive description of two of these assumptions, the ones that characterize
inequity aversion (the other two assumptions are made for
mathematical convenience). One assumption is that, keeping the
relative payoff $\sigma_i$ constant, $i$'s utility is increasing in
$x_i$. Thus, for two choices that give the same relative share, 
player $i$'s decision is consistent with payoff maximization. The
second assumptions is that, holding $x_i$ constant, $i$'s utility is
strictly concave in $\sigma_i$, with a maximum at the point at which
player $i$'s monetary payoff is equal to the average
payoff. Thus, keeping monetary payoff constant, players 
prefer an equal distribution of monetary payoffs. This
utility function was 
shown to fit empirical data quite well in a number of games,
including the 
dictator game, 
ultimatum game,
and 
prisoner's dilemma.
(We defer a comparison of Bolton and Ockenfels' approach with that of
Fehr and Schmidt.)

Shortly after the explosion of inequity-aversion models, several
economists observed that some decision-makers appear to act in a way
that \emph{increases} inequity, if 
this
increase 
results in an increase in the total payoff of the participants
\citep{charness2001relative,kritikos2001distributional,andreoni2002giving,charness2002understanding}. This
observation is hard to reconcile with inequity-aversion models, and
suggests that people not only prefer to minimize
inequity, but also prefer to maximize social welfare. 

To estimate these preferences, \citet{andreoni2002giving} 
conducted an experiment, found the utility function in a particular
class of utility functions that best fit 
the experimental
results, and 
showed that this utility function also fits
data from other experiments well. In more detail, they conducted
an experiment in which participants made decisions in a series of
modified dictator games where the cost of giving is in the set
$\{0.25, 0.5, 1, 2, 3\}$. For example, when the cost of giving is
$0.25$, sending one token to the recipient results in the recipient
receiving four tokens. Andreoni and Miller found
that 22.7\% of the dictators were perfectly selfish (so their
behavior could be rationalized by the utility function
$u(x_1,x_2)=x_1$), 14.2\% of dictators split the monetary payoff
equally with the recipient (so their behavior could be
rationalized by the \emph{Rawlsian} utility function
$u(x_1,x_2)=\min(x_1,x_2)$,%
\footnote{Named after John Rawls, a philosopher, who argued, roughly
 speaking, that in a just society, the social system should be
 designed so as to maximize the payoff of those worst off.}
 and 6.2\% of the dictators gave to the
recipient only when the price of giving was smaller than 1 (and thus
their behavior could be rationalized by the \emph{utilitarian} utility
function $u(x_1,x_2)=\frac{1}{2} x_1+ \frac{1}{2} x_2$). To
rationalize the behavior of the 
remaining 57\% of the dictators, Andreoni and Miller fit
their data to a utility function of the form
$$
u_1(x_1,x_2)=\left(\alpha x_1^\rho+(1-\alpha)x_2^\rho\right)^{1/\rho}.
$$
Here $\alpha$ represents the extent to which the dictator (player 1)
cares about his own monetary payoff more than
that of the
recipient (player 2), while $\rho$ represents the convexity of
preferences. Andreoni and Miller found that subjects can be divided into
three classes that they called \emph{weakly selfish} ($\alpha=0.758,
\rho=0.621$; note that if $\alpha = 1$, we get
self-regarding
preferences), \emph{weakly Rawlsian} ($\alpha=0.654,
\rho=-0.350$; note that if $0 < \alpha < 1$ and 
$x_1, x_2 > 0$,
then
as $\rho \rightarrow -\infty$, we converge to Rawlsian preferences), and
\emph{weakly utilitarian}  
($\alpha=0.576, \rho=0.669$; note that if $\alpha=.5$ and $\rho = 1$,
we get utilitarian preferences). Moreover, they showed that the model
also fits well experimental results on the standard dictator game, 
the
public-goods game, and 
prisoner's dilemma. Finally, this model can also
explain the results mentioned above showing that some decision-makers act so as
to increase inequity, if the increase leads 
to an increase in social welfare.

\citet{charness2002understanding} considered a more general family of utility
functions
that includes the ones mentioned earlier as special
cases. Like Andreoni and Miller, they conducted an experiment to
estimate the parameters of the 
utility function that best fit the data. For simplicity, we describe
Charness and Rabin's
utility
%
function in the case of two players; we refer to their paper for the
general formulation. They considered a utility function for player 2 of the form
\begin{equation*}\begin{split}
    u_2(x_1,x_2)=&(\rho r+\sigma s)x_1+\\
&(1-\rho r - \sigma s)x_2,
\end{split}\end{equation*}
%
where: (1) $r=1$ if $x_2>x_1$, and $r=0$ otherwise; (2) $s=1$ if
$x_2<x_1$, and $s=0$ otherwise.\footnote{The general form of the
  utility function has a third component that takes into account reciprocity in
    sequential games. Since in this review we focus on
  normal-form games, we ignore this component here.}
Intuitively, $\rho$ represents how important it is to agent 2 that he
gets a higher payoff than agent 1, while $\sigma$ represents how
important it is to agent 2 that agent 1 gets a higher payoff than he
does.  
Charness and Rabin did not make any \emph{a priori} assumptions regarding
$\rho$ and $\sigma$. But they showed that 
by setting $\rho$ and $\sigma$ appropriately, 
one
can
recover the earlier models: 
\begin{itemize}
\item Assume that $\sigma\leq\rho\leq0$. In this case, 
player 2's utility is increasing in 
$x_2-x_1$. 
So, by definition, this case
      corresponds to competitive preferences.
    \item Assume that $\sigma<0<\frac12<\rho<1$.
      If  $x_1<x_2$, then 
            player 2's utility is $\rho x_1+(1-\rho)x_2$, and thus
            depends positively on both $x_1$ and  $x_2$, because $0<\rho<1$.
    Moreover, since $\rho>\frac12$, player 2 prefers
            increasing player 1's payoff to his own;
       that is, player 2 prefers to decrease inequity (since $x_2 > x_1$).  
  If $x_2<x_1$, then player 2's utility is $\sigma x_1+(1-\sigma)x_2$,
  and thus depends 
negatively on $x_1$
and positively on $x_2$, so again players 2 prefers to decrease inequity.
\item If $0<\sigma\leq\rho\leq1$, then player 2's utility depends
positively on both $x_1$ and $x_2$. 
Charness and Rabin define these as \emph{social-welfare preferences}. (Note
that in the case $\sigma=\rho=\frac12$, one obtains utilitarian
preferences $u_2(x_1,x_2)=(x_1+x_2)/2$; more generally, Charness and
Rabin apply the term ``social-welfare preferences'' to those
preferences where the 
individual payoffs are both 
weighted
positively.)  
\end{itemize}

To test which of these cases better fits the experimental data,
Charness and Rabin
conducted a series of dictator game
experiments. They found that the standard assumption of narrow
self-interest explains only 68\% of the data, assuming 
competitive preferences
(i.e., $\sigma\leq\rho\leq0$) 
explains even less (60\%), and that assuming
a preference for inequity aversion 
(i.e., $\sigma<0<\frac12<\rho<1$)
is consistent with 75\% of
the data.  The best results were obtained by assuming a preference for
maximizing social welfare (i.e., $0<\sigma\leq\rho\leq1$), 
which explains 97\% of the data.

\citet{engelmann2004inequality} also showed that assuming a preference
for maximizing social welfare leads to better predictions than
assuming a preference for inequity aversion.  They considered a set of
decision problems designed to compare the relative ability of several
classes of utility functions to make predictions.  They considered
utility functions corresponding
to
payoff maximization,
social-welfare maximization, Rawlsian maximin preferences, 
the utility function proposed by Fehr and
Schimdt%
, 
and
the utility functions proposed by
Bolton and Ockenfels. They found that the best fit of the data was
provided by a combination of social-welfare concerns, maximin preferences,
and selfishness. Moreover, 
Fehr and Schimdt's
inequity-aversion model outperformed that of Bolton and Ockenfels.
However, this increase in performance was entirely driven by
the fact that, in many cases, the predictions of Fehr and
Schimdt 
reduced
to maximin preferences.

\emph{Team-reasoning} models
\citep{gilbert1987modelling,bacharach1999interactive,sugden2000team}
and equilibrium notions such as \emph{cooperative equilibrium}
\citep{HR1,capraro2013model} also take social-welfare maximization
seriously.   The underlying idea of these approaches is that
individuals do not always act so as to maximize their individual
monetary payoff, but may also take into account social welfare.
For example, in the
prisoner's dilemma, social welfare is maximized by mutual
cooperation, so team-reasoning models predict that people cooperate. 
However, 
since these models typically assume that the utility of a player 
is a function of the sum of the payoffs of all players,
they cannot explain 
behaviors in zero-sum games, such as the
 dictator 
game.  Another
limitations of these approaches is their inability to explain the
behavior of people who choose to 
minimize
both their individual
payoffs and social welfare, such as responders who reject
offers in the ultimatum game.

\cite{cappelen2007pluralism} introduced a model in which participants
strive to balance their individual ``fairness ideal'' with their
self-interest. They consider three
fairness ideals.
\emph{Strict egalitarianism} contends that people are not
responsible for their effort and talent; according to this view,
the fairest distribution of resources is the equal
distribution.
\emph{Libertarianism} argues that people are responsible
for their effort and talent; according to this view, resources should be
shared so that each person's share is in proportion to what
s/he produces.
\emph{Liberal egalitarianism} is based on the belief
that people are responsible for their effort, but not for their talent;
according to this view, resources should be distributed so as
to minimize differences due to talent, but not those due to effort. To
formalize people's tendency to implement their fairness ideal,
Cappelen et al.~considered utility functions of the form
$$
u_i(x_i,a,q)=\gamma_i x_i - \frac{\beta_i}{2}\left(x_i-p_i(a,q)\right)^2,
$$
where $a=(a_1,\ldots,a_n)$ is the vector of talents of the players,
$q=(q_1,\ldots,q_n)$ is the vector of efforts made by the players,
$x_i$ is the monetary payoff of player $i$, and $p_i(a,q)$ is the
monetary payoff that player $i$ believes to be his 
fair 
share. Finally, $\gamma_i$ and $\beta_i$ are
individual parameters, representing the extent to which player $i$
cares about his monetary payoff and his fairness ideal. In order
to estimate the distribution of the fairness ideals,
Cappelen et al.~conducted an experiment using a variant
of the dictator game that includes a production phase. The quantity produced
depends on factors within the participants' control (effort) and
factors beyond participants' control (talent). The 
experiment showed that 39.7\% of the subjects can be viewed as
strict egalitarians, 43.4\% as
liberal egalitarians, and 16.8\% as libertarians. Although this approach
is useful in cases where the initial endowments are earned by the
players through a task involving effort and/or talent, when the
endowments are received as windfall gains, as in most laboratory
experiments, it reduces to inequity aversion, and so
it suffers from the same limitations that other utility
functions based on inequity aversion do.

The frameworks discussed thus far are particularly suitable for studying
situations in which there is only one decision-maker or in which the
choices of different decision-makers are made simultaneously. In many
cases, however, 
there is more than one decision maker, and they make their
choices 
sequentially.
For non-simultaneous
move games, scholars have long recognized that 
intentions
play an important role. 
A particular class of intention-based models---reciprocity models---is 
based on the idea that
people typically reciprocate 
(perceived)
good actions with
good actions and 
(perceived)
bad actions with bad actions. 
Economists have considered several 
models of reciprocity
\citep{rabin1993incorporating,levine1998modeling,charness2002understanding,dufwenberg2004theory,sobel2005interdependent,falk2006theory}. 
Intention-based models have been shown to explain deviations
from outcome-based predictions in a number of situations 
where beliefs about the other players' intentions may play a role
\citep{falk2003nature,mccabe2003positive,fehr2006economics,falk2008testing,dhaene2010sequential}.  
Although
we acknowledge the existence and the importance of these models, as we
mentioned in the introduction, in this review we focus on 
normal-form games. Some of these games (e.g., dictator game, die-under-cup paradigm, trade-off game) 
have only one decision maker, and therefore beliefs about others'
intentions play no role.  
In this contexts, beliefs can still play a role, 
for example
beliefs about others' beliefs, since even in single-agent games, 
there
may be others watching (or the decision maker can play as if
there are).
This leads us to psychological
games, which will be mentioned in Section \ref{se:language_based}.

Outcome-based social preferences explain many experimental results well. In
particular, looking at the experimental regularities listed in 
Section~\ref{sec:regularities}, they easily
explain the first three regularities, at least qualitatively. However,
they cannot explain any of the remaining regularities. Indeed, the
main limitation of outcome-based preferences is that 
they depend only on the monetary
payoffs.

As we suggested in Section~\ref{sec:regularities}, one way of explaining
these regularities is to assume that people have
\emph{moral preferences}.
Crucially, these moral preferences cannot be
defined solely in terms of the economic consequences of the available
actions.
We review such moral preferences
in Section \ref{se:moral_pref}.%
\footnote{
Of course, we can consider 
outcome-based preferences in the context of decision rules other than
expected-utility maximization,
such as \emph{maximin}
expected utility \citep{Wald50,GS82,GS1989}%
, 
\emph{minimax regret}
\citep{Niehans,Savage51}, 
and
\emph{maximin} expected utility
  \citep{Wald50,GS82,GS1989}.
 For example, with maximin, an agent chooses the act that maximizes her
worst-case 
(expected)
utility.  With minimax regret,
an agent chooses the act that minimizes her worst-case regret (the gap between the payoff for an act $a$ in state $s$ and
the payoff for the best act in state $s$). 
Although useful in many contexts, none of these approaches
can explain Experimental Regularities 4-7,
since they are all outcome-based.
}

Before moving to moral preferences, it is worth noting that another limitation of outcome-based social preferences is that they predict
correlations between different behaviors that are not consistent with
experimental data.  
For example, \cite{chapman2018econographics} 
reported on
a large experiment showing that eight standard measures
of prosociality can actually be clustered in 
three principal components, corresponding to altruistic behavior, punishment, and inequity aversion, which are virtually unrelated.
Chapman et al.~observed that this is not consistent with any
outcome-based social preferences. 
However, we will see that this result is consistent 
with moral preferences.

\section{Moral preferences}\label{se:moral_pref}

In the previous sections, we showed that outcome-based preferences
and, more generally, outcome-based decision rules, are inconsistent
with Experimental Regularities 4-7.
In this section, we review the empirical literature suggesting
that all seven experimental regularities discussed
in Section \ref{sec:regularities} can be explained by assuming that
people have 
preferences for following 
a
norm,%
\footnote{We wrote ``\emph{a} norm'', and not ``\emph{the} norm'', because
there are different types of norms. For the 
aim of this review, it is important to distinguish between
personal beliefs about what is right and wrong, beliefs about what
 others approve or disapprove of, and beliefs about what others actually
 do. We will get to this distinction in more details later.}
 and
discuss attempts to formalize this using an appropriate utility function.
We use the term \emph{moral preferences} as an umbrella term
to denote this type of utility function.

\emph{Experimental Regularities 1-4.} The fact that donations in the
standard
dictator game, offers and rejections in the ultimatum game,
cooperation in social dilemmas, and honesty in lying tasks can be
explained by moral 
preferences was independently shown by many
authors. 

\cite{krupka2013identifying} asked dictators to report, for
each available choice, how ``socially appropriate'' they think other
dictators think that choice is; dictators were incentivized to guess
the modal answer of other dictators. They found
that an equal split was rated as the most socially appropriate
choice. They also found that framing effects in the dictator
game when passing from the ``give'' frame to the ``take'' frame can be
explained by a change in the perception of what the most
appropriate action is.

\cite{kimbrough2016norms} introduced a
``rule-following'' task to measure people's \emph{norm-sensitivity},
specifically, how important it was to them to follow
rules of behavior. In this task, each participant has to control a
stick figure walking across the screen from left to right. Along its
walk, the figure encounters five traffic lights, each of which turns
red 
when the figure approaches it.
Each participant has
to decide how long to wait at the traffic light (which turns green
after five seconds), knowing that s/he will lose a certain amount of
money for each second spent on the waiting. The total amount of time
spent waiting is taken as an individual measure of
norm-sensitivity. Kimbrough and Vostroknutov found that this parameter
predicts giving in the dictator game and the public-goods game,
and correlates with rejection thresholds in the ultimatum game.
\cite{bicchieri2010behaving} showed that ultimatum-game responders
reject the same offer at different rates, depending on the other
available offers; in particular, responders tend to accept offers that
they consider to be fairer, compared to the other available offers. 

In a similar vein, \cite{capraro2018right} and \cite{capraro2019power}
found that giving in the dictator game depends on what people perceive to be
the morally right thing to do. 
Indirect evidence that moral preferences drive giving in the dictator game
was also provided by research showing that including moral
reminders in the instructions of the dictator game increases giving
\citep{branas2007promoting,capraro2019increasing}.
In addition, as mentioned in Section \ref{sec:regularities},
\cite{eriksson2017costly} showed that framing
effects among responders in the ultimatum game can be explained by a change
in the perception of what is the morally right thing to do, 
while \cite{dal2014right} found that moral reminders increase cooperation in
the iterated prisioner's dilemma. 


We remark that \cite{capraro2018social} found that, in the 
ultimatum
game, 92\% of the proposers and 72\% of responders declare that
offering half is the morally right thing to do, while
\cite{capraro2018right}
found that 81\% of the subjects declare that
cooperating is the morally right thing to do in the one-shot prisoner's
dilemma.


Finally, the
fact that honest behavior in economic games in which participants can
lie for their benefit is partly driven by moral preferences was
suggested by several authors
\citep{gneezy2005deception,erat2012white,fischbacher2013lies,abeler2019preferences}. 
Empirical
evidence was provided by 
\cite{cappelen2013we},
who found that telling the truth in
the sender-receiver game in the Pareto white-lie condition correlates
positively with giving in the dictator game, suggesting that `aversion to lying not only is positively associated 
with pro-social preferences, but for many a stronger moral motive than the concern for the welfare of others'. 
\cite{biziou2015does} replicated the correlation between honesty in the Pareto white-lie condition
and giving in the dictator game and additionally showed a similar correlation 
with cooperation in the prisoner's dilemma; the authors suggested that 
cooperating, giving, and truth-telling might be driven by a common motivation
to do the right thing.
Finally, \cite{bott2017you} found that moral reminders
decrease tax evasion in a field experiment with Norwegian tax-payers. 

\emph{Experimental Regularities 5-6.} \cite{list2007interpretation}%
, \cite{bardsley2008dictator}, and \cite{cappelen2013give}
showed that people tend to be more
altruistic in the standard dictator game than in the dictator
game with a take option (Experimental Regularity 5). The fact that
this behavioral change might reflect moral preferences was
suggested by \cite{krupka2013identifying}.
They found that sharing nothing in the
standard dictator game is considered to be far less socially
appropriate than sharing nothing in the dictator game with a take
option. They also showed that
social appropriateness can explain why some dictator-game givers
prefer to avoid the interaction altogether, given the chance to do
so (Experimental Regularity 6): dictators rate exiting the game to be
far less socially inappropriate than keeping the money in the standard
dictator game. 

\emph{Experimental Regularity 7.} In Section~\ref{sec:regularities},
we reviewed the literature showing that behavior in several
games, including the dictator game, the prisoner's
dilemma, the ultimatum game, and the trade-off game, depends on the
language used to present the instructions, especially when it
activates moral concerns. 

To summarize, all seven regularities can be qualitatively explained by
assuming that people have moral preferences. In what follows, we
review the models of moral preferences that have been
introduced thus far in the literature.
The idea that morality has
to be incorporated in economic models has been around since the
foundational work of Adam Smith and Francis Y. Edgeworth
\citep{smith2010theory,edgeworth1881mathematical}; see
\citep{sen1977rational,binmore1994game,tabellini2008institutions,bicchieri2006grammar,enke2019kinship}
for more recent accounts.
However, work on utility functions that
take moral preferences into account is relatively recent.  
In this review, we focus on utility functions that can be
applied to all or to most%
\footnote{Economists have also introduced a number of domain-specific
models; for example, models to explain cooperation in the prisoner's dilemma
\citep{bolle1990prisoners}, honesty in lying tasks
\citep{abeler2019preferences, gneezy2018lying}, fairness in
principal-agent models \citep{ellingsen2008pride}, and honesty in principal-agent models
\citep{alger2006screening,alger2007screening}.
Although useful in their contexts, these models cannot be readily
extended to other types of interaction. 
} of the economic interactions
that are 
described in terms of 
normal-form games with monetary payoffs.%
\footnote{Economists have also studied models of morality in other
games (see, e.g., \cite{benabou2011identity}. Recently, economists have also sought to explain political behavior in terms of moral preferences
\citep{bonomi2021identity,enke2022morals}. Although these models cannot 
be readily applied to the economic games that are the focus of this review, they show
that the idea that moral preferences can help explain people's behavior
is gaining traction across different areas of research.}


We proceed chronologically. We start by discussing the work of
\cite{akerlof2000economics}. 
Their motivation was to study ``gender discrimination
in the workplace, the economics of poverty and social exclusion, and
the household division of labor''. To do so, they proposed a utility
function that takes into account a person's \emph{identity}. The
identity is assumed to carry information about how a person should
behave, which, in turn, is assumed to depend on the social categories
to which a person belongs. 
In this setting, Akerlof and Kranton considered a
utility function of the form 
$$
u_i=u_i(a_i,a_{-i},I_i),
$$
where $I_i$ represents $i$'s identity%
.
They showed that their model can qualitatively
explain group differences such as the ones that motivated their
work. This model is certainly
consistent with Experimental Regularities 1-7. Indeed, it suffices
to assume that the identity takes into account a tendency to 
follow the norms.
This model is conceptually similar to a previous model proposed by
\cite{stigler1977gustibus}, 
which is based on the idea that preferences should not be defined over
marketed goods,  
but over general commodities that people transform into consumption goods. Although
\cite{stigler1977gustibus} do not aim to explain the experimental regularities that are 
the focus of this review, their model is consistent with 
them: for 
example, people may cooperate in order to maintain good relations, or may act altruistically 
to experience the warm glow of giving, or may act morally to adhere to their self-image. 
We refer to \cite{sobel2005interdependent} for a more detailed
discussion and for the mathematical equivalence between
this model and that of Akerlof and Kranton.


A more specific model, but one based on a similar idea, was
introduced by \cite{brekke2003economic}. Their initial aim was to
explain field experiments showing that paying people to
provide a public good can ``crowd out'' intrinsic motivations. For
example, paying donors 
reduces
blood donation
\citep{titmuss2018gift}, 
people's willingness to accept a
nuclear-waste repository in their neighborhood
\citep{frey1997cost}, and 
volunteering 
\citep{gneezy2000pay}.
To explain these findings,
Brekke, Kverndokk, and Nyborg 
considered economic interactions in which each player $i$
($i=1,\ldots,n$) has to put in some effort $e_i$, measured in units of
time, to generate a public good $g_i$. 
At most $T$ units of time are 
assumed
to be available,
so that $i$ has to
decide how much time 
$e_i$
to contribute to the public good and how
much time $l_i$ to use for leisure: 
$l_i = T - e_i$. 
The total quantity of
public good is $G=G_p+\sum_{i=1}^n g_i$, where
$G_p$ is the public provision of the public good. The monetary payoff
that player $i$ receives from putting in effort $e_i$ is denoted 
$x_i$. The key assumption of the model is that player $i$ has a
morally ideal effort, denoted $e_i^*$. 
Brekke, Kverndokk, and Nyborg 
postulated
that player $i$ maximizes a
utility function of the form
$$
u_i=u_i(x_i,l_i)+v_i(G) + f_i(e_i,e_i^*),
$$
where $u_i$ and  $v_i$ are increasing and concave,
while the function $f(e_i,e_i^*)$ 
is assumed to attain its
maximum at $e_i=e_i^*$; we can think of $f$ as taking into account the
distance between $i$'s actual effort $e_i$ and the ideal effort
$e_i^*$ in an inversely related way. 
As an explicit example, Brekke, Kverndokk, and Nyborg 
considered
the function
$f_i(e_i,e_i^*)=-a(e_i-e_i^*)^2$, with $a>0$.
Therefore, \emph{ceteris paribus},
players aim 
to maximize their monetary payoff, their leisure time, 
and
the public
good, while aiming at minimizing the distance from their moral
ideal. 
Brekke, Kverndokk, and Nyborg 
supposed
that, before deciding their action,
players consider their morally ideal effort. They 
assumed
that all players share a utilitarian moral philosophy, so that the morally ideal effort
is found by maximizing $W=\sum_iu_i$ with respect to $e_i$. 
Under these assumptions, they showed
that their 
model
is consistent with the crowding-out effect.
Specifically, they showed that when a fee is introduced for people who
do not contribute 
to the public good, if this fee is at least equal to the cost of
buying $g_i$ units of public good in the market 
and if this fee is smaller than the utility corresponding to the gain
of leisure time due to not contributing, 
then the moral ideal $e_i^*$ is equal to $0$; in other words, the
fee becomes a moral justification for 
not contributing. This intuitively happens because individuals leave
the responsibility of ensuring the public good 
to the organization: the public good is provided by the organization,
which buys it using the fees;  this is convenient 
 for the individuals,
as they gain in leisure time.
If we replace time with money, then
this utility function can capture the empirical
regularities observed in the dictator game and social dilemmas. However,
the utility function cannot easily be applied to settings
that do not have the form of a public-goods game, 
such as the 
ultimatum 
and trade-off games. 

A more general utility function was introduced by
\cite{benabou2006incentives}. It tries to take into account
altruism and is motivated by a theory of social signalling,
according to which people's actions are associated with reputational
costs and benefits
\citep{NS,smith2000turtle,gintis2001costly}.  Players
choose a participation level
for some prosocial activity
from an action set $A\subseteq\mathbb R$.
Choosing $a$ has a cost $c(a)$ and gives a monetary payoff
$ya$; $y$ can be positive, negative, or zero. Players are assumed to
be characterized by
a type
$v=(v_a,v_y)\in\mathbb R^2$, where $v_a$
is, roughly speaking, the 
impact of 
the prosocial factors associated with
participation level $a$ on the agent's utility,
while $v_y$ is, roughly speaking, the 
impact of money on the agent's utility.
B\'enabou and Tirole 
mentioned
that $v_a$, the
utility
of choosing $a$, is determined by at least
two factors: the material payoff of the
other player and the enjoyment derived from the act of giving.
Thus, their 
approach
can capture Andreoni's
(\citeyear{andreoni1990impure}) notion of warm-glow giving.

B\'enabou and Tirole then 
defined
the \emph{direct benefit} of
action $a$ to be 
$$
D(a)=(v_a+v_yy)a-c(a),
$$
and the \emph{reputational benefit} to be
$$
R(a)=x[\gamma_aE(v_a|a,y)-\gamma_yE(v_y|a,y)],
$$
where $E(v|a,y)$ represents the observers' posterior expectation that
the player is of type $v$, given that the player chose $a$
when the monetary incentive to choose $a$ is $y$. 
The parameters
$\gamma_a$ and $\gamma_y$ are assumed to be non-negative%
. To understand this assumption, note that, by definition, players
with high $v_a$ are 
prosocial;
players with high $v_y$ are
greedy. Therefore, the hypothesis that $\gamma_a$ and $\gamma_y$ are
non-negative formalizes the idea that people like to be perceived as
prosocial 
($\gamma_a\geq0$) and not greedy ($\gamma_y\geq0$).
 The parameter 
 $x>0$ represents the visibility of an action, that is, the probability
that the action is visible to other players.
B\'enabou and Tirole then defined the utility function  
$$
u(a)=D(a)+R(a).
$$
They studied this utility function in a number of contexts in which
actions can be observed and decision options can be defined in terms
of contribution.
While useful in its domains of applicability, this utility
function cannot be applied to games where
choices cannot be described in terms of participation levels, 
such as trade-off games in which players have the role of distributing
money, without themselves being affected by the distribution levels. 

A utility function that captures moral motivations and can be
applied to all games was introduced by \cite{levitt2007laboratory}. 
They 
argued
that the utility of player $i$ when
s/he chooses action $a$ depends on two factors%
. The first factor is
the utility corresponding to the monetary payoff associated with action $a$. It
is 
assumed 
to be increasing in the
monetary value of $a$, denoted $x_i$%
. The second factor is
the moral cost or
benefit $m_i$ associated with $a$. 
Levitt and List
focused on three factors
that can affect the moral value of an action. The first is the
negative externality that $a$ imposes on other people. They
hypothesized that the externality is an increasing function of $x_i$:
the more a player receives from choosing $a$, the less other
participants receive. The second factor is the set $n$ of moral norms
and rules that govern behavior in the society in which the decision-maker
lives. For example, the very fact that an action is illegal may
impose an additional cost for that behavior.
The third factor is the extent to which actions are observed. For example, if
an illegal or an immoral
action is recorded, or performed in front of the experimenter, it is
likely that the decision-maker pays a greater moral cost than if 
the same action is performed when no one is watching. The effect
of scrutiny is denoted by $s$; greater scrutiny is assumed to increase
the moral cost. Levitt and List added this component to
take into account 
the fact that, when the
behavior of participants cannot be monitored, people tend to be less
prosocial than when it can
\citep{bandiera2005social,list2006behavioralist,benz2008people}.
They proposed that player $i$ maximizes the utility function 
$$
u_i(a,x_i,n,s) = m_i(a,x_i,n,s) + w_i(a,x_i).
$$
This approach can explain Experimental Regularities 1, 3,
5, and 7.
However, the situation described in Experimental Regularity 2
and some instances of Experimental Regularities 4 and 6 violate Levitt
and List's assumptions. Specifically, the assumption that the 
the negative externalities associated with player $i$'s action depend
negatively on $i$'s monetary payoff does not hold in the
ultimatum game, where rejecting a low offer (the choice which,
is typically viewed as the moral choice \citep{bicchieri2010behaving})
decreases 
both players' monetary payoff.  This is also the case in the sender-receiver
game in the Pareto white-lie condition, where 
telling the truth is viewed as the moral choice, yet minimizes the
monetary payoffs of both
players. But these are minor limitations; they can be addressed by
considering a slightly more general utility function that
depends not only on $x_i$, but also on $\sum_{j\neq i}x_{j}$ (and
dropping the assumption that these two variables are inversely related). 

A similar approach was used by \cite{lopez2008aversion},
although he focused on extensive-form games. He
assumed the existence of a ``norm'' function $\Psi_i$ that
associates with each information set $h$ for player $i$ in a given game an
action $\Psi_i(h)$ that can be taken at $h$. Intuitively,
$\Psi_i(h)$ is the moral choice at information set
$h$.  L\'opez-P\'erez assumed that player $i$ receives a
psychological disutility (in the form of a negative emotion, such as
guilt or shame) whenever he violates the norm expressed by
$\Psi_i$.
He did not propose a specific functional expression
for the utility function; rather, he studied a particular example of a
norm, the \emph{E-norm}. This norm is
conceptually similar to the one used by
\cite{charness2002understanding}, described in Section
\ref{se:social_preferences}. L\'opez-P\'erez showed that
a utility function that gives higher utility to strategies
that perform more moral actions 
qualitatively fits the empirical data in several
contexts. Although he did not explicitly show that the experimental
regularities presented in Section \ref{sec:regularities} can be
explained using his model, it is easy to show that this is 
the case. Indeed, it suffices to 
assume 
that the mapping $\Psi_i$
just associates with the game the morally right action for player $i$.
(We can view 
a  normal-form game as having a single information set, so we can view
$\Psi_i$ as just applying to the whole game.)
Note, however, that the norm $\Psi_i$ defined in this way is different from the
E-norm considered by L\'opez-P\'erez, which is outcome-based.

\cite{andreoni2009social} focused on the dictator game and introduced
a utility function that combines elements from theories of social
image with inequity aversion. The fact that some people care
about how others see them has been recognized by economists for at
least two decades \citep{bernheim1994theory,glazer1996signaling}.
Combining these ideas with inequity aversion,
Andreoni and Bernheim proposed that dictators maximize the utility
function 
$$
u_i(x_i,m,t)=f(x_i,m)+tg\left(x_i-\frac12\right),
$$
where $x_i\in[0,1]$ is the monetary payoff of the dictator (the
endowment is normalized to 1), $m\geq0$ is the social image of the
dictator, $f$ represents the utility associated with $x_i$
(which is 
assumed 
to be increasing in both $x_i$ and $m$ and concave in $x_i$),
$t\geq0$ is a parameter representing the extent to which the dictator
cares about minimizing inequity, and $g$ is a (twice continuously
differentiable, strictly concave) function that attains its maximum at
$0$, that formalizes the intuition that the dictator gets a
disutility from not implementing the equal distribution. 
Andreoni and Bernheim assumed that there
is an audience $A$ that includes the recipient and
possibly other people (e.g., the experimenter). The audience observes
the donation $1-x_i$ and then forms beliefs about the dictator's level
of fairness $t$.
They 
assumed
that the dictator's social image is some function $B$ of
$\Phi$, the cumulative distribution $\Phi$ representing
$A$'s beliefs about the dictator's level of fairness.
For example, the function $B$ could be the mean
of $t$ given
$\Phi$. Andreoni and Bernheim showed that, with some minimal
assumptions about $\Phi$, we can explain donations in the dictator
game quite
well. In particular, their utility function is consistent with the
prevalence of exactly equal splits, that is, the lack of donations 
slightly above or slightly below 50\% of the endowment. They
also considered a variant of the dictator game in which with
some probability $p$ the dictator's choice is not implemented, but
instead the recipient receives an amount $x_0$ close to 0. Intuitively,
this should have the effect of creating a peak of voluntary donations at
$x_0$, since dictators can excuse the outcome as being
beyond their control, thus preserving their social image.  Andreoni
and Bernheim observed this behavior, and
showed that this is indeed consistent with their utility function.
Unfortunately, it is not clear how to extend this utility function beyond dictator-like games.

A conceptually similar approach was considered by
\cite{dellavigna2012testing}. Instead of formalizing a concern for
social image, their utility function takes into account the effect of social
pressure (which might affect people's decisions through social
image). This approach was motivated by a door-to-door campaign with three
treatments: a \emph{flier} treatment, in which households were
informed one day in advance by a flier on their doorknob of the
upcoming visit of someone soliciting donations; a \emph{flier with an
  opt-out checkbox} 
treatment, where the flier contained a box to be checked in case the
household did not want to be disturbed; and a \emph{baseline}, in which
households were not informed about the upcoming
visit. Della Vigna, List, and Malmendier found that the flier decreased the
frequency of 
the door being opened, compared to the baseline. Moreover, the flier with
an opt-out check box also decreased giving, but the effect was
significant only for small donations. To explain these findings,
they considered a two-stage game between a \emph{prospect} (potential
donor) and a 
solicitor. In the first stage, the  prospect may receive a flier and, if
so, he notices the flier with probability $r\in(0,1]$. In the second
stage, the solicitor visits the home. The probability of the
prospect opening the door is denoted by $h$: if the prospect did
not notice the flier, $h$ is equal to the baseline probability $h_0$;
otherwise, the prospect can update this probability at a cost $c(h)$,
with $c(h_0)=0$, $c'(h_0)=0$, $c'' > 0$. That is, not updating the
probability of opening the door has no cost; updating it has a cost
that depends monotonically on the adjustment. A donor can donate
either 
in person or through other channels
(e.g., by email). Let $g$ be the amount donated in person and $g_m$ the
amount donated through other means. A donor's utility is taken to be
\begin{equation*}\begin{split}
u(g,g_m)=&f(w-g-g_m)+\\
&\alpha v(g+\delta g_m,g_{-i})+\\
&\sigma(g^\sigma-g)\mathbf 1_{g<g^\sigma},
\end{split}\end{equation*}
where $w$ represents the initial wealth of the donor;
$f(w-g-g_m)$ represents the utility of private consumption; $\alpha$
is a parameter representing the extent to which the donor cares about
the payoff of the charity, which can be negative;\footnote{In fact, if
  the social pressure (third addend of the utility function) is high
  enough, someone can end up donating even though 
he
dislikes the
  charity.} $\delta$ is the proportion of the donation made through
other channels that does not reach the intended recipient (e.g., the cost of 
an envelope and a stamp); $g_{-i}$ is the (expected) donation made by
other donors; $\sigma$ is a parameter representing the extent to which
the donor cares about social pressure; $g^\sigma$ is a
trade-off donation: if the donor donates less than $g^\sigma$ when the
solicitor is present, then the donor pays a cost depending on
$g^\sigma-g$.  Della Vigna, List, and Malmendier showed that this utility
function captures their experimental results well. However,
it is not clear how to extend it to 
qualitatively different decision contexts. 

\cite{kessler2012norms} considered a utility function in which players receive a
disutility when they deviate from a norm;
this leads to a model similar 
to
that of \cite{brekke2003economic}
that we described above. 
The main difference is that, instead of considering time, \cite{kessler2012norms} 
considered
money. 
Players are assumed to have a norm $\widehat x$ that
represents the ideal monetary contribution. The utility of player $i$
%
is defined as 
\begin{equation*}\begin{split}
u_i(x_i,x_j,\widehat x)=&\pi_i(x_i,x_j)-\\
&\phi_ig(\widehat x-x_i)\mathbf1_{x_i<\widehat x},
\end{split}\end{equation*}
%
where $\pi_i(x_i,x_j)$ is the monetary payoff of player $i$ when
$i$ contributes $x_i$ and $j$ contributes $x_j$,
$\phi_i$ is a parameter representing $i$'s norm-sensitivity,
and $g$ represents the disutility
that player $i$ gets from deviating from the norm, so $g(x_i) = 0$ if
$x_i \ge \widehat x$ and $g(x_i)$ increases with $x_i - \widehat{x}$
if $x_i < \widehat{x}$. 
Kessler and Leider applied their utility function to an experiment
involving four games: an additive two-player public-goods game,
a multiplicative public-goods game,
a double-dictator game,
and a
Bertrand game.
The set of contributions available
depended on the game; they were chosen to ensure 
a mismatch between the individual
monetary payoff-maximizing action and the socially beneficial
action. Participants played ten rounds of these games, with random
re-matching. Some of these rounds were preceded by a contracting phase
in which participants could agree on which contribution to
choose. Kessler and Leider observed that the presence of a contracting phase
increased contributions. They argued that their utility function fits
their data well; in particular, contracting increased
contribution by increasing the norm.

In subsequent work,
\citet{kimbrough2016norms} used this approach to explain prosocial
behavior in the public-goods game, the dictator game, and the ultimatum
game. The key innovation of this work is the estimation of the
parameter $\phi_i$, which was done using the ``rule-following
task'' task discussed earlier in this
section. They found that their measure of
norm-sensitivity significantly correlates with cooperation in the
public-goods game, with giving in the dictator game, and with
rejection thresholds in the ultimatum game (but not with 
offers in the ultimatum game, although results 
trend
in the expected
direction). In sum, this utility function is very useful in its domain
of applicability. Moreover, although it might seem difficult to extend
it to situations in which strategies (and norms) cannot be expressed
in terms of contributions, it can easily be extended to games where the space of strategy profiles is a 
metric space
$(X,d)$
where $d$ is the metric (i.e., a distance function between strategies).
In this setting, we
can replace $\widehat x - x$ in the utility function with
$d(\widehat x,x)$. It is easy to see that this utility function can
explain all seven experimental regularities, assuming that $\widehat
x$ coincides with what people view as the moral
choice. 

\cite{lazear2012sorting} considered a utility function motivated by their
empirical finding that some sharers in the 
dictator game prefer to avoid 
the
dictator-game interaction, given the possibility of doing so. 
They considered situations where players can choose 
one of two
scenarios. In the scenario with a sharing option, the agent
plays the standard dictator game in the role of the dictator: 
he
receives an endowment $e$, which 
he
has to split between 
himself
($x_1$) and the recipient ($x_2$). In the scenario without a sharing
option, the player simply receives $x_1=e$, while the recipient
receives $x_2=0$.
They proposed a utility that depends
on the scenario: $u=u(D,x_1,x_2)$, where $D=1$ represents the
scenario with a sharing opportunity, while $D=0$ represents the
scenario without the sharing opportunity. The fact that the utility
depends on the scenario makes it possible to classify people
as one of  three types. The first type, the ``willing
sharers'', consists of individuals who prefer the sharing
scenario and to share  their endowment (at least, to some
extent). Formally, this 
type of player is characterized by the conditions
$$
\begin{cases}
  \text{max}_{x_1\in[0,e]}u(1,x_1,e-x_1) > u(0,e,0)\\
  \text{argmax}_{x_1\in[0,e]}u(1,x_1,e-x_1) < e.
\end{cases}
$$
The second type, the ``non-sharers'', never share. They are
defined by the condition
$$
\text{argmax}_{x_1\in[0,e]}u(1,x_1,e-x_1) = e.
$$
The third type, called  ``reluctant sharers'', are perhaps most
interesting. They are determined by the
remaining conditions: 
$$
\begin{cases}
    \text{max}_{x_1\in[0,e]}u(1,x_1,e-x_1) < u(0,e,0)\\
\text{argmax}_{x_1\in[0,e]}u(1,x_1,e-x_1) < e.
\end{cases}
$$
The first condition says that these players prefer to avoid the
sharing opportunity when given the possibility of doing so. 
However, if they are forced to play the dictator game,
they share part of their
endowment.  Lazear, Malmendier, and Weber showed 
that there are a significant number of people of the third type.  
Indeed, some subjects even prefer to pay a cost to avoid the
sharing opportunity. While this gives a great deal of insight, 
it is hard to generalize this type of utility function to other kinds
of interaction.  

\cite{krupka2013identifying} introduced a model in which subjects are
torn between following their self-interest and following the
``injunctive norm'', that is, what they believe other people would
approve or disapprove of \citep{cialdini1990focus}. Let
$A=\{a_1,\ldots,a_k\}$ be the set of actions available.
Krupka and Weber assumed the existence of a
social norm 
function
that associates to each action $a_j \in A$ a number
$N(a_j)\in\mathbb R$ representing the degree of social appropriateness
of $a_j$. $N$ is 
hypothesized
to be independent of the individual; it
represents the extent to which society views $a_j$ as socially
appropriate. $N(a_j)$ can be negative, in which case $a_j$ is
viewed as socially inappropriate. The utility of
player $i$ is defined as 
$$
u_i(a_j)=v_i(\pi_i(a_j))+\gamma_i N(a_j),
$$
where $\pi_i(a_j)$ is the monetary payoff of player $i$ associated with
action $a_j$, $v_i$ is the utility associated to monetary payoffs, and
$\gamma_i$ is the extent to which $i$ cares about doing what is
socially appropriate. As we mentioned earlier in this section, one of
Krupka and Weber's contributions was to introduce a
 method of measuring social appropriateness.
People were shown the available actions and, for each of them,
asked to rate how socially appropriate they were.
Participants were also
incentivized to guess the mode of the choices made by the other
participants.
It is not difficult to show that Krupka and Weber's
utility function is consistent 
with all the experimental regularities, at least if we assume
that the
most socially appropriate choice coincides with what people
believe to be the morally right thing to do. This suggests that one
possible limitation of Krupka and Weber's approach
is that it takes into account
only the injunctive norm. In general, there are situations 
where what people believe others would approve of (the
injunctive norm) is different from the choice they believe to be
morally right (the personal norm). For example, suppose that a vegan 
must decide whether to buy a steak for \$5 or a vegan meal
for \$10.  Knowing that the vast majority of the population eats
meat, the vegan might believe that others would view the steak as the most
socially appropriate choice. However, since the vegan
thinks that eating meat is morally wrong, 
she
would opt
for buying the vegan meal. This suggests that there might be
situations where social appropriateness
might not be a good predictor of human
behavior. We return to this issue in Section \ref{se:outlook}. 

\cite{alger2013homo} introduced a notion of \emph{homo moralis}. They
considered a symmetric 2-player game where the players
have a common strategy set $A$.\footnote{Their results also apply
  to non-symmetric games where players do not know \emph{a priori}
  which role they will play.}  
Let $\pi_i$ be $i$'s payoff function; since the game is symmetric, we have
that $\pi_i(a_i,a_j) = \pi_j(a_j,a_i)$.  Players may differ in how
much they care 
about morality. Morality is defined by taking inspiration from Kant's
\emph{categorical imperative}: one should make the choice that is
\emph{universalizable}, that is, the action that would be best if
everyone took it (e.g., cooperating in the prisoner's dilemma).%
\footnote{See \cite{laffont1975macroeconomic} for a
  macroeconomic application of this principle.
 \cite{roemer2010kantian} defined a notion of \emph{Kantian
   equilibrium} in public-goods type games; these are strategy profiles
in  which no player would prefer all other players to change their
  contribution levels by the same multiplicative factor.} More
specifically, they defined player $i$ to be a \emph{homo moralis} if
his
utility function has the form
$$
u_i(a_1,a_2)=(1-k)\pi_i(a_1,a_2)+k\pi_i(a_1,a_1),
$$
where $k\in[0,1]$ represents the degree of morality. If
$k=0$, then we recover the standard \emph{homo economicus}; if
$k=1$, we have \emph{homo kantiensis}, someone who makes the
universalizable choice.
Alger and Weibull proved that 
evolution
results in 
the degree of morality $k$ being equal to the
\emph{index of assortativity} of the matching process.
We refer to their paper for the exact definition of the index of
assortativity. 
For our purposes, all that is relevant is that it is a non-negative
number that takes into 
account the potential non-randomness of the matching process: it is zero
with random matching, and greater than zero for non-random matching
processes. 
Thus, in the particular case of random matching, Alger and Weibul's
theorem shows that evolution results in homo moralis with degree of
morality $k=0$, namely,
homo economicus. 
If the matching is not random,
assortativity can favor the emergence of
homo moralis
with a degree of morality strictly greater than zero.

In subsequent work, \cite{alger2016evolution} showed
that homo moralis preferences are evolutionary stable, according to 
an appropriate definition of evolutionary stability for
preferences.
This approach is certainly useful for understanding the
evolution of morality. However, if the payoff function
$\pi$ is equal to the monetary payoff, then homo moralis preferences are outcome-based, so cannot explain Experimental Regularities 4-7. On the other hand, if the payoff function is not equal to the monetary payoff (or otherwise outcome-based), then how should it be defined?
Despite these limitations, it is important to note that there are some
framing effects that can be explained by  
the model of
\cite{alger2013homo}. Suppose, for example, that market interactions
are formed through a  
matching process with less assortativity than are business partnership interactions, or co-authorship interactions, 
then 
a lower
degree of morality should be expected in market interactions. 

Kimbrough and Vostroknutov
(\citeyear{kimbrough2020injunctive,kimbrough2020theory}) proposed a
utility function 
similar to the one proposed by Krupka
and Weber (\citeyear{krupka2013identifying}) that was discussed
earlier. Specifically, they proposed that people are torn between
maximizing their material payoff and (not) doing what they think
%
%
society would (dis)approve of. This corresponds to the utility function
$$u_i(a)=v_i(\pi_i(a))+\phi_i\eta(a),$$
where $v_i(\pi_i(a))$ represents the utility from the monetary payoff
$\pi_i(a)$ corresponding to action $a$, $\phi_i$ represents the extent
to which player $i$ cares about (not) doing what he thinks society
would (dis)approve of, and $\eta(a)$ is a measure of the extent to which
society approves or disapproves $a$. The key difference between this and
Krupka and Weber's approach is in the definition of $\eta$, which
corresponds to Krupka and Weber's $N$. While
Krupka and Weber defined $N$ empirically, by asking
experimental subjects to guess what they believe others would find
socially (in)appropriate, Kimbrough and Vostroknutov
defined $\eta$ in terms of 
the cumulative dissatisfaction
that players
experience when a certain strategy profile is realized, rather than
other possible strategy profiles.
Kimbrough and Vostroknutov's notion of ``dissatifaction'' corresponds
to what is more typically called \emph{regret}: 
the difference between what a
player could have gotten and what he actually got. Thus, the
cumulative dissastisfaction is 
the sum of the regrets of the
players.  
In their model, Kimbrough and Vostroknutov 
assumed
that the
normatively best outcome is the one that minimizes aggregated
dissatisfaction. 
This implies that this model
predicts that Pareto dominant strategy profiles are always more
socially appropriate than Pareto dominated strategy
profiles. Therefore, while this model explains well some types of
moral behaviors, such as cooperation, it fails to explain moral
behaviors that are Pareto dominated, such as honesty when lying is
Pareto optimal and framing effects in the trade-off game that push
people to choose the equal but Pareto dominated allocation of money. 

\cite{capraro2021mathematical} introduced a utility function which,
instead of  
considering people's tendency to follow the injunctive norm,
considers their tendency to follow their personal norms, that is, their internal standards about what is right or wrong 
in a given situation \citep{schwartz1977normative}. Specifically,
Capraro and Perc  
proposed the utility function
$$u_i(a)=v_i(\pi_i(a))+\mu_iP_i(a),$$
where $v_i(\pi_i(a))$ represents the utility from the monetary payoff
$\pi_i(a)$ corresponding to action $a$, $\mu_i$ represents the extent
to which player $i$ cares about doing what he thinks to be morally right, and
$P_i(a)$ represents the extent to which player $i$ thinks that $a$ is morally right.
From a practical perspective, the main difference between
this utility function and 
those proposed by \cite{krupka2013identifying} and by \cite{kimbrough2016norms} is that
personal norms are individual, 
that is, $P_i(a)$ depends specifically on player $i$, while the
injunctive norm depends
only on the society where an individual lives, and so is the same
for all individuals in the same society. 
This utility function is also consistent with all seven regularities
described in Section \ref{sec:regularities}. 

Finally, \cite{bavsic2021personal} proposed a utility function that combines 
personal and injunctive norms:
$$u_i(a)=v_i(\pi_i(a))+\gamma_i S(a) + \delta_i P_i(a),$$
where $\gamma$ and $\delta$ represent the extent to which people care about
following the injunctive and the personal norms, respectively. This utility function was suggested
to the authors by experimental evidence in support of the fact that injunctive norms and personal norms 
are differentially associated with giving in the dictator game (standard version and a variant
with a tax) and with behavior in the ultimatum game (as well as in a third-party punishment game). 
Also this utility function is consistent with all the seven experimental regularities that we 
are considering in this review.

At the end of Section \ref{se:social_preferences}, we observed that
another limitation of social preferences is 
that they predict correlations between different prosocial behaviors that
are not observed in experimental data. We conclude this section
by observing that the lack of these correlations is consistent with at least some moral
preferences. Indeed, several models of moral preference (e.g.,
\cite{akerlof2000economics,levitt2007laboratory,lopez2008aversion,krupka2013identifying,kimbrough2016norms,capraro2021mathematical,bavsic2021personal})
do not assume that morality is unidimensional.
If morality is multidimensional and  
different people may weigh different dimensions differently, then 
it is possible that, for some people, 
the right thing to do is to act altruistically, while for others it is to
punish antisocial behavior, and for yet  
others, it is to minimize inequity; this would rationalize the
experimental findings of \cite{chapman2018econographics}.  
We will come back to
the multidimensionality of morality in Section \ref{se:outlook}.

\section{Language-based preferences}\label{se:language_based}
In this section, we go beyond moral preferences and consider settings
where an agent's utility depends, not just on what happens, but on
how the agent feels about what happens, which is largely captured by the
agent's beliefs, and how what happens is described, which is
captured by the agent's language.

Work  on a formal model, called a \emph{psychological game}, for taking an 
agent's beliefs into account  in the utility function
 started with
\cite{GPS}, and was later extended by \cite{BD09} to allow for dynamic
beliefs, among other things.  The overview by \cite{BD20} shows how the
fact that the utility function in psychological games can 
depend on beliefs allows us to capture, for example, 
guilt
feelings (whether Alice leaves a tip for a taxi driver Bob in a
foreign country might depend on how guilty Ann would feel if she
didn't leave a tip, which in turn depends on what Alice believes Bob
is expecting in the way of a tip),   
reciprocity (how kind player $i$ is to $j$  depends on $i$'s beliefs
regarding whether $j$ will reciprocate), other 
emotions (disappointment, frustration, anger, regret), image (your belief about
how others perceive you), and expectations (how what you actually get
compares to your expectation of what you will get; see also
\citep{KR06}), among other things.

Having the agent's utility function depend on \emph{language}, how the
world is described, provides
a yet more general way to express preferences.  (It is more general
since the language can include a description of the agent's beliefs;
see below.)
Experimental Regularity 7 describes several cases in which people's
behavior depends on the words used to describe the available actions. 
Language-based preferences allow us to explain Experimental Regularity 7,  
as well as all the other experimental regularities regarding social
interactions  
listed in Section 2, in a straightforward way. From this point of view, we can 
interpret language-based models as a generalization of moral preferences, 
where the utility of a sentence is assumed to carry the moral value of
the action  
described by that sentence. 
Language-based models are strictly more general than moral preferences. 
In this section, we will show that they
can explain other well-known regularities (e.g., ones not involving
social interactions) 
that have been found in behavioral experiments, such as the Allais paradox.

A 
classic example 
is 
standard
\emph{framing
effects}%
, where people's decisions depend on whether the
alternatives 
are described in terms of gains or losses \citep{tversky1985framing}.
  It is well known, for example, that presenting alternative medical
treatments in terms of survival rates versus mortality rates can produce
a marked difference in how those treatments are evaluated, even by
experienced physicians \citep{MPST82}.
A core insight of \emph{prospect theory}
\citep{KT79}
is
that subjective value depends not (only) on facts about the
world, but on how those facts are viewed (as gains or
losses, dominated or undominated options, etc.).
And how they are viewed
often depends on how they are described in language. For example,
\cite{Thaler80} 
observed
that the credit card lobby preferred that the
difference between the price charged to cash and credit card customers
be presented as a discount for paying cash rather than as a surcharge for
paying by credit card.  The two different 
descriptions amount to taking different reference points.%
\footnote{Tversky and Kahneman emphasized the
distinction between gains and losses in prospect theory, but they
clearly understood that other features of a description were also
relevant. However, note that
prospect theory applied to monetary outcomes results is yet another instance
of outcome-based preferences, so cannot explain Experimental
Regularities 4--7.} 

Such language dependence is ubiquitous.  
We celebrate 10th and 100th anniversaries specially, and make a big deal
when the Dow Jones Industrial Average crosses a multiple of 1,000, all
because we happen to work in a base 10 number system.
Prices often end in .99, since people seem to perceive differently the
difference between \$19.99 and \$20 and the difference between \$19.98
and \$19.99.  
We refer to
\citep{Shlain19} for some recent work on and an
overview of this well-researched topic.

One important side effect of the use of language,  to which we return
below, is that it emphasizes categories and clustering.  
For example,
as \cite{KC94} showed, when people were asked to estimate the average
high and low temperatures in Providence, Rhode Island, on various
dates, while they were fairly accurate, their estimates were relatively
constant for any given month, and then jumped when the month
changed---the difference in estimates for two equally spaced days was
significantly higher if the dates were in different months than if
they were in the same month.  
This clustering arises in likelihood estimation as well.   We often assess
likelihoods using words like ``probable'', ``unlikely'', or ``negligible'',
rather than numeric representations; and when numbers are used, we
tend to round them \citep{MM10}.


The importance of language to economics was already stressed by
Rubinstein
in his book \emph{Economics and Language} 
\citep{Rubinstein00}. 
In Chapter 4, for example, he considers the
impact of having an agent's preferences be definable in a simple
propositional language.
There have been various formal models that take
language into account.  For example:
\begin{itemize}
\item \citet{Lip99}
consider ``pieces of information'' that an agent might
receive, and takes the agent's state space to be
characterized by maximal pieces of information.  He also applies his
approach to framing problems, among other things.
  \item Although
\citet{AE07} do not consider language explicitly, they do allow for
possibility that there may be different descriptions of a particular
event, and use this possibility to capture framing.  For them, a
``description'' is a partition of the state space.
\item \cite{BEH06} take an agent's  object of choice to be
programs, where a program is either a primitive program or has the form
{\bf if} $t$ {\bf then} $a$ {\bf else} $b$, where $a$ and $b$ are
themselves programs, and $t$ is a test.  A test is just a
propositional formula, so language plays a significant role in the
agent's preferences.  Blume, Easley, and Halpern too show how framing
effects can be captured in their approach.
\item There are several approaches to decision-making that 
can be viewed as 
  implicitly based on language.  For example, a critical component of
  Gilboa and Schmeidler's \emph{case-based} 
decision theory \citep{GS01} is the notion of a \emph{similarity function},
  which assesses how close a pair of problems are to each other.  We can
  think of problems as descriptions of choice situations in some language.
  Jehiel's notion of \emph{analogy-based expectation
  equilibrium} \citep{Jeh05} assumes that there is some way of
  partitioning situations in bundles that, roughly speaking, are treated
  the same way when it 
comes
to deciding how to move in a game.  Again,
  we can think of as these bundles as ones whose descriptions are
  similar.  Finally, \citet{Mull02} assumes that people use coarse
  categories (similar in spirit to Jehiel's analogy bundles, 
  the categories partition the space of possibilities) to make
  predictions.  While none of these approaches directly models the
  language used, many of the examples they use are language-based.
  
\item While not directly part of the utility function, the role of
  vagueness and ambiguity 
  in language, how it affects communication, and its
  economic implications have been studied and modeled
(see, e.g.  \citep{BO14,HK15}).
\end{itemize}

We focus here on \emph{language-based games} \citep{bjorndahl2013language},
where the utility function directly depends on the language.
As we shall see, language-based games provide a way of formalizing all
the examples above.  
The following example, which deals with surprise,
gives a sense of how language-based games work.


\begin{example} \emph{\citep{bjorndahl2013language}}  \label{exa:spp}
\emph{Alice and Bob have been dating for a while now, and Bob has decided
that the time is right to pop the big question. Though he is not one
for fancy proposals, he does want it to be a surprise. In fact, if
Alice expects the proposal, Bob would prefer to postpone it entirely
until such time as it might be a surprise. Otherwise, if Alice is not
expecting it, Bob's preference is to take the opportunity.}

\emph{We can summarize this scenario by the payoffs for Bob given in
Table~\ref{tbl:spp}.}
\begin{table}[h]
\center
\begin{tabular}{r|c|c}
& $p$ & $\lnot p$\\
\hline
$B_{A} \, p$ & 0 & 1\\
\hline
$\lnot B_{A} \, p$ & 1 & 0\\
\end{tabular}
\caption{The surprise proposal.} \label{tbl:spp}
\end{table}

\noindent \emph{In this table, we denote Bob's two strategies, proposing and not
proposing, by $p$ and $\neg p$, respectively, and use $B_A p$ (respectively,
$\neg B_A p$) to denote that Alice is expecting (respectively, not expecting)
the  proposal.  (More precisely, $B_A p$ says that Alice believes that
Bob will propose; we are capturing Alice's expectations by her beliefs.)
Thus, although Bob is
the only one who moves in this game, his utility depends, not just on
his moves, but on Alice's expectations.}
\bbox
\end{example}

This choice of language already illustrates one of the
features of the language-based approach: coarseness.  We
used quite a coarse language to describe Alice's
expectation: she either expects the proposal or she doesn't.
Since the expectation is modeled using belief, this example can be
captured using a psychological game as well.
Of course, whether or not Alice expects a proposal may be
more than a
binary affair: she may, for example, consider a proposal unlikely,
somewhat likely, very likely, or certain.
In a psychological game, Alice's beliefs would be expressed 
by placing an arbitrary probability $\alpha \in [0,1]$ on $p$.  But
there is good reason to think that an accurate model of her expectations
involves only a small number $k$ of distinct
``levels'' of belief, rather than a continuum. Table \ref{tbl:spp}, for
simplicity, assumes that $k=2$, though this is easily generalized to
larger values.

Once we fix a language (which is just a finite or infinite set of
formulas), we can take a \emph{situation} to be a maximal consistent
set of formulas; that is, a complete description of the world in that
language.%
\footnote{What counts as a maximal consistent set of formulas depends
  on the semantics of the language.    We omit the (quite standard)
  formal details here; they can be found in \citep{bjorndahl2013language}.}
In the example above, there are four situations: $\{p,
B_Ap\}$ (Bob proposes and 
Alice expects the proposal), $\{ p, \neg B_A
p\}$ (Bob proposes but Alice is not expecting it), $\{\neg p, B_A p\}$
(Bob does not propose although Alice is expecting him to), and $\{\neg
p, \neg B_A p\}$ (Bob does not propose and Alice is not expecting a
proposal).  An agent's language describes all the features of the game that
are relevant to the player.  An agent's utility function associates a
utility with each situation, as in Table \ref{tbl:spp} above.
Standard game theory is the special case where,
given a set $\Sigma_i$ of strategies (moves) for each player $i$, the
formulas have the form $\play_{i}(\sigma_{i})$ for $\sigma_i \in
\Sigma_i$.  The situations are then strategy profiles.

A normal-form psychological game can be viewed as a special case of
a language-based game where (a) the language talks only about agent's
strategies and agents' possibly higher-order beliefs about these
strategies (e.g., Alice's beliefs about Bob's
beliefs about Alice's beliefs about the proposal), and (b) those
beliefs are described using probabilities.  For example, taking
$\alpha$ to denote Alice's probability of $p$, psychological game
theory might take Bob's utility function to be the following:
$$
u_{B}(x, \alpha) = \left\{ \begin{array}{ll}
1 - \alpha & \textrm{if $x = p$}\\
\alpha & \textrm{if $x = \lnot p$.}
\end{array} \right.
$$
The function $u_{B}$ agrees with Table \ref{tbl:spp} at its
extreme points if we identify $B_{A} p$ with $\alpha = 1$ and $\lnot B_{A} p$
with $\alpha = 0$. Otherwise, for the continuum of other values that $\alpha$
may take between 0 and 1, $u_{B}$ yields a 
convex
combination of the
corresponding extreme points. Thus, in a sense, $u_{B}$ is a
continuous approximation to a scenario that is essentially
discrete.

The language implicitly used in psychological games is rich
in one sense---it allows a continuum of possible beliefs---but is poor
in the sense that it talks only about belief.
That said, as
we mentioned above, many human emotions can be expressed naturally
using beliefs, and thus studied in the context of psychological games.
The following example  illustrates
how.

\begin{example} \label{exa:ina} \emph{\citep{bjorndahl2013language}}
\emph{Alice and Bob play a classic prisoner's dilemma game, with one
twist: neither wishes to live up to low expectations. Specifically, if
Bob expects the worst of Alice (i.e., expects her to defect), then Alice, indignant at Bob's opinion of her, prefers to
cooperate. Likewise for Bob. On the other hand, in the absence of such low expectations from their opponent, each will revert to their
classical preferences. }

\emph{The standard prisoner's dilemma is summarized in Table \ref{tbl:cpd}:}
\begin{table}[h]
\center
\begin{tabular}{c|c|c}
& \textsf{c} & \textsf{d}\\
\hline
\textsf{c} & (3,3) & (0,5)\\
\hline
\textsf{d} & (5,0) & (1,1)\\
\end{tabular}
\caption{The classical prisoner's dilemma.} \label{tbl:cpd}
\end{table}

\emph{Let $u_{A}$, $u_{B}$ denote the two players' utility functions according
to this table.  Let the language consist of the formulas of the form
$\play_i(\sigma)$, $B_i(\play_i(\sigma))$, and their negations,
where $i \in \{A,B\}$ and 
$\sigma \in \{\textsf{c},\textsf{d}\}$. 
 Given a situation $S$,
let $\sigma_S$ denote the unique strategy profile determined by $S$.
We can now define a
language-based game that captures the intuitions above by taking
Alice's utility function $u_A'$ on situations $S$ to be
\begin{displaymath}
u_{A}'(S) = \left\{ \begin{array}{ll}
u_A(\sigma_S) - 6 &
\mbox{if $\play_{A}(\textsf{d})$,}\\
&B_{B} \play_{A}(\textsf{d}) \in S\\
u_A(\sigma_S)& \textrm{otherwise,}
\end{array} \right.
\end{displaymath}
and similarly for $u_{B}'$.}

\emph{More generally, we could take take Alice's utility to be
$u_A(\sigma_S) - 6\theta$ if $\play_{A}(\textsf{d}), B_{B}
\play_{A}(\textsf{d}) \in S$, where $\theta$ is a measure of the
extent to which Alice's indignance affects her utility.  And yet more
generally, if the language lets us talk about the full range of
probabilities, Alice's utility can depend on the probability she
ascribes to $\play_{A}(\textsf{d})$.
(Although we have described the last variant using language-based games,
it can be directly expressed using psychological games.)}
\bbox
\end{example}

Using language lets us go beyond expressing the belief-dependence
captured by psychological games.  For one thing, the coarseness of the
utility language lets us capture some well-known anomalies in
the preferences of consumers.
For example, we can formalize the
explanation hinted at earlier for why prices often end in .99.
Consider a language that consists of price ranges like ``between \$55 and
\$55.99'' and ``between \$60 and \$64.99''. With such a language, the
agent is forced to ascribe the same utility to \$59.98 and \$59.99,
while there can be a significant difference between the utilities of
\$59.99 and \$60.  Intuitively, we think of the agent as using
two languages: the (typically quite rich) language used to describe the world
and the (perhaps much coarser) language over which utility is
defined.  Thus, while the agent perfectly well understands the
difference between a price of \$59.98 and \$59.99, her utility
function may be insensitive to that difference.

Using a coarse language effectively limits the set of describable
outcomes, and thus makes it easier for a computationally bounded agent
to determine her own utilities. These concerns suggest that there
might be even more coarseness at
higher ranges.  For example,
suppose that the language includes terms like ``around \$20,000'' and
``around \$300''.  
If we assume that both ``around \$20,000'' and ``around \$300''
describe intervals (centered at \$20,000 and \$300, respectively), it
seems reasonable to assume that the interval described by ``around
\$20,000'' is larger than that described by ``around \$300''.
Moreover, it seems reasonable that \$19,950 should be in the first
interval, while \$250 is not in the second.
With this choice of language (and the further assumptions), we can capture
consumers who might drive an extra 5 kilometers to
save \$50 on a \$300 purchase but would not be willing to drive an
extra 5 kilometers to save \$50 on a \$20,000 purchase (this point was
already make by \cite{Thaler80}):
a consumer gets the same utility if they pay \$20,000 or \$19,950 (since
in both case, they are paying ``around \$20,000), but do not get the
same utility paying \$250 rather than \$300.

This can be viewed as an application of \emph{Weber's law}, 
which asserts that the minimum difference between two stimuli
necessary for a subject to discriminate between them
is proportional to the magnitude of the stimuli; thus, larger stimuli
require larger differences between them to be perceived.  
Although traditionally applied to physical stimuli, Weber's law
has also been
shown to be applicable in the realm of numerical perception: larger
numbers are subjectively harder to discriminate from one another
\citep{ML67,Restle}.

As we observed earlier, 
we can understand
the partitions that arise in
Jehiel's notion of 
a coarsening of the language; this is even more explicit in
Mullainathan's notion of categories.
The observation of \cite{MM10} that people often 
represent likelihoods using words suggests that coarseness can arise
in the representation of likelihood.  To see the potential impact of
this on decision-theoretic concerns, 
consider the following analysis of \emph{Allais' paradox}
\citep{Allais53}.
\begin{example} \label{exa:all}
\emph{Consider the two pairs of gambles described in
Table \ref{tbl:all}.}
\begin{table}[h]
\center
\begin{tabular}{l l | l l}
\multicolumn{2}{c|}{Gamble 1a} & \multicolumn{2}{c}{Gamble 1b}\\ \hline
$1$ & \$1 million & $.89$ & \$1 million\\
 & & $.1$ & \$5 million\\
 & & $.01$ & \$0\\
\multicolumn{4}{c}{}\\
\multicolumn{2}{c|}{Gamble 2a} & \multicolumn{2}{c}{Gamble 2b}\\ \hline
$.89$ & \$0 & $.9$ & \$0\\
$.11$ & \$1 million & $.1$ & \$5 million
\end{tabular}
\caption{The Allais paradox.} \label{tbl:all}
\end{table}

\noindent \emph{The first pair is a choice between (1a) \$1 million for sure, versus
(1b) a $.89$ chance of \$1 million, a $.1$ chance of \$5 million, and a
$.01$ chance of nothing. The second is a choice between (2a) a $.89$
chance of nothing and a $.11$ chance of \$1 million, versus (2b) a $.9$
chance of nothing and a $.1$ chance of \$5 million. The ``paradox''
arises from the fact that most people choose (1a) over (1b), and most
people choose (2b) over (2a) \citep{Allais53}, but these preferences are not
simultaneously compatible with expected-utility maximization. }

\emph{Suppose that we apply the observations of \cite{MM10} to this
setting.  Specifically, 
suppose that probability judgements such as ``there is a $.11$ chance
of getting \$1 million'' are represented in a language with only
finitely many levels of likelihood. In particular, suppose that the
language has only the
descriptions ``no chance'', ``slight chance'', ``unlikely'', and their
respective opposites, ``certain'', ``near certain'', and ``likely'',
interpreted as in Table \ref{tbl:cla}.}
\begin{table}[h]
\center
\begin{tabular}{c|c|c}
Range & Description & Representative\\
\hline
$1$ & certain & $1$\\
$[.95,1)$ & near certain & $.975$\\
$[.85,.95)$ & likely & $.9$\\
$(.05,.15]$ & unlikely & $.1$\\
$(0,.05]$ & slight chance & $.025$\\
$0$ & no chance & $0$\\
\end{tabular}
\caption{Using coarse likelihood.} \label{tbl:cla}
\end{table}

\emph{Once we represent likehoods using words in a language rather than
numbers, we have to decide how to determine (expected) utility.
For definiteness, suppose that the utility of a gamble as described in this
language is determined using the interval-midpoint representative
given in the third column of Table \ref{tbl:cla}. Thus, a ``slight
chance'' is effectively treated as a $.025$ probability, a ``likely''
event as a $.9$ probability, and so on.}

\emph{Revisiting the gambles associated with the Allais paradox,
suppose that we replace the actual probability given in
Table~\ref{tbl:all} by the word that represents it (i.e., replace 1 by
``certain'', .89 by ``likely'', and so on)---this is how we assume
that an agent might represent what he hears.  Then when doing an
expected utility calculation, the word is replaced by the probability
representing that word, giving us Table~\ref{tbl:ala}.
\commentout{
\begin{table}[h]
\center
\begin{tabular}{r l | r l}
\multicolumn{2}{c|}{Gamble 1a} & \multicolumn{2}{c}{Gamble 1b}\\ \hline
certain & \$1 million & likely & \$1 million\\
 & & unlikely & \$5 million\\
 & & slight chance & \$0\\
\multicolumn{4}{c}{}\\
\multicolumn{2}{c|}{Gamble 2a} & \multicolumn{2}{c}{Gamble 2b}\\ \hline
likely & \$0 & likely & \$0\\
unlikely & \$1 million & unlikely & \$5 million
\end{tabular}
\caption{The Allais paradox, coarsely described.} \label{tbl:alc}
\end{table}
For one thing, probabilities of $.89$ and $.9$ are not distinguished
at all (nor are $.1$ and $.11$), which immediately implies that (2b)
is preferred to (2a), provided $u_{A}(\$5\textrm{ million}) >
u_{A}(\$1\textrm{ million})$. On the other hand, likelihoods of $0$
and $.01$ are not only distinguished by this language, but their
difference is exaggerated. Table \ref{tbl:ala} shows the result of
substituting the approximations from Table \ref{tbl:cla} in for the
descriptions of Table \ref{tbl:alc}. }
\begin{table}[h]
\center
\begin{tabular}{l l | l l}
\multicolumn{2}{c|}{Gamble 1a} & \multicolumn{2}{c}{Gamble 1b}\\ \hline
$1$ & \$1 million & $.9$ & \$1 million\\
 & & $.1$ & \$5 million\\
 & & $.025$ & \$0\\
\multicolumn{4}{c}{}\\
\multicolumn{2}{c|}{Gamble 2a} & \multicolumn{2}{c}{Gamble 2b}\\ \hline
$.9$ & \$0 & $.9$ & \$0\\
$.1$ & \$1 million & $.1$ & \$5 million
\end{tabular}
\caption{The Allais paradox, coarsely approximated.} \label{tbl:ala}
\end{table}
}

\emph{Using these numbers, we can calculate the revised utility of (1b) to be
$.9 \cdot u_{A}(\$1\textrm{ million}) + .1 \cdot u_{A}(\$5\textrm{
  million}) + .025 \cdot u_{A}(\$0),$
and this quantity may well be less than $u_{A}(\$1\textrm{ million})$,
depending on the utility function $u_{A}$. For example, if
$u_{A}(\$1\textrm{ million}) = 1$, $u_{A}(\$5\textrm{ million}) = 3$,
and $u_{A}(\$0) = -10$, then the utility of gamble (1b) evaluates to
$.95$. In this case, Alice prefers (2b) to (2a) but also prefers (1a)
to (1b).  Thus, this choice of language rationalizes the observed
preferences of many decision-makers.  (\cite{Rubinstein00} offered
a closely related analysis.)}

\emph{It is worth noting that this approach to evaluating gambles will lead to
discontinuities; the 
utility
of a gamble that gets, say%
,
\$1,000,000
with probability $x$ and \$5,000,000 with probability $1-x$ does not
converge to the 
utility
of a gamble that gets \$1,000,000 with
probability 1 as $x$ approaches 1.  Indeed, we would get
discontinuities at the boundaries of every range.  We expect almost
everyone to treat certainty specially, and so have a special category
for the range $[1,1]$; what people take as the range for other
descriptions will vary.  \cite{AS09} present an approach to the Allais
paradox that is based on the discontinuity of the utility of gambles
at 1, and present experimental evidence for such a discontinuity.   We
can view the language-based approach as providing a potential
explanation for this discontinuity.}
\bbox
\end{example}

Going back to Example~\ref{exa:ina}, note that cooperating is rational
for Alice if she thinks that Bob is sure that she will defect, since
cooperating in this case would yield a minimum utility of 0, whereas
defecting would result in a utility of $-1$. On
the other hand, if Alice thinks that Bob is \emph{not} sure that she will
defect, then since her utility in this case is determined
classically, it is rational for her to defect, as usual.
\cite{bjorndahl2013language} define a natural generalization of
Nash equilibrium in language-based games and show that, in
general---and, in particular in this game---%
it does
not exist, even if
mixed strategies are allowed.  The problem
is the discontinuity in payoffs.  Intuitively,
a Nash equilibrium is a state of play where players are
happy with their choice of strategies \emph{given correct beliefs about what their opponents will choose}. But there is a fundamental tension
between a state of play where everyone has correct beliefs, and
one where some player successfully surprises another.

\cite{bjorndahl2013language} also define a natural generalization of
the solution concept of \emph{rationalizability}
\citep{Ber84,Pearce84}, and show that all 
language-based games where the language satisfies a natural
constraint have rationalizable strategies.  But the question of
finding appropriate solution concepts for language-based games remains
open.
Moreover, the analysis of \cite{bjorndahl2013language} was carried
out only for normal-form games.
\cite{GPS} and \cite{BD09} consider extensive-form psychological games.
Extending language-based games to the extensive-form setting will
require dealing with issues like the impact of the language changing
over time.  

We conclude this section by observing that, interpreting the choice of
language as a framing of the game, language-based games can
be seen as a special case of framing.
There have already been attempts to provide general models of
the effects of framing.
For example, \cite{tversky1993context} considered situations in which
an agent’s choices  
may depend on the background set $B$ (i.e., the set of all available choices)  
and the choice set $S$ (i.e., the set of offered choices). Tversky and
Simonson introduced 
 a choice function $V_B(x,C) = v(x) + \beta f_B(x) + \theta g(x,S)$
consisting of three components:  
 $v(x)$ is the context-free value of $x$, independent of $B$, $f_B(x)$
captures the effect of the background,  and
$g(x,S)$ captures the effect of the choice set.
 \cite{salant2008f} and 
\cite{ellingsen2012social} assumed 
that there is a set $\mathcal F$ of frames and that the utility function 
depends on the specific frame $F\in\mathcal F$.  

%
These models of framing effects can easily explain all seven regularities
by choosing suitable frames and utility functions.
For example,
Ellingsen et al.
applied
their model to the prisoner's dilemma
and were able to explain changes in the rate
of cooperation depending on the name of the game (`community game'
vs. `stock market game'), 
under the assumption that
the frame affects the 
beliefs about the opponent's level of altruism.
Although these models can be applied to explain framing effects
specifically generated by language, 
they do not model the effect of language directly.
As
Experimental Regularity 7 shows, 
many framing effects are in fact ultimately due to language. 
Language-based games provide a 
way of capturing
these language effects directly.
Moreover, they
allow us to ask questions that are not asked in the standard framing
literature, such as,  
for example, why people's behavior changes when the price of gas goes from 
\$3.99 to \$4.00, but not when it goes from \$3.98 to \$3.99. (This would not typically 
be called a framing effect; but we can reinterpret it as a framing effect
by assuming that there is a frame $F\in\mathcal F$ such that
``over \$4'' and ``under \$4''  
are different categories in $F$.)

\section{Future research and outlook}\label{se:outlook}

The key takeaway message of this article is that the monetary payoffs 
associated with actions are not sufficient to fully explain 
people's behavior. What matters are not just the monetary payoffs, but also 
the language used to present the available actions. It follows that 
economic models of behavior should also take language into account. 
We believe that the shift from outcome-based preferences to
language-based preferences will have a profound impact on economics. We conclude our review with a discussion of some lines of research that we believe will play a prominent role in this shift. These lines of research are quite interdisciplinary, involving psychology, sociology, philosophy, and computer science.

In the previous sections, we have highlighted experimental results
suggesting that, at least  in some cases,
people seem to have moral preferences. However, we were
deliberately vague about where these moral preferences come from. 
Do they arise from 
personal beliefs about what is right and wrong,  beliefs about what
others
approve or disapprove of, or beliefs about what others actually
 do? 

Moral psychologists and moral philosophers have long
argued that there are several types of norms, which sometimes 
conflict with one other. An important distinction is between
\emph{personal norms} and \emph{social norms}
\citep{schwartz1977normative}. Personal norms refer to internal
standards about what is considered to be right and wrong; 
they are not externally motivated by the approval or disapproval of
others. It might happen that others either approve or disapprove of them, 
but this is not what drives the personal norms. Social norms, on
the other hand, refer to ``rules and standards that are understood by
members of a group, and that guide and/or constrain behavior without
the force of laws'' \citep{cialdini1998social}. 
Two important types of social norms are 
\emph{injunctive norms} and 
\emph{descriptive
  norms}, defining, respectively, what people think others would 
approve or disapprove of and what people actually do
\citep{cialdini1990focus}.
A unified theory of norms has been proposed 
more recently by Cristina Bicchieri
(\citeyear{bicchieri2006grammar}).
According to her theory, there are
three main classes of norms: \emph{personal normative beliefs}, which
are personal beliefs about what should happen in a given situation;
\emph{empirical expectations}, which are personal beliefs about what one expects others to do; and \emph{normative expectations}, which are personal beliefs about what others think one should do. 

Although the different types of norms often align, as we discussed earlier, they may conflict. When descriptive norms
conflict with injunctive norms%
, people tend to follow the
descriptive norm, as shown by a famous field experiment in which
people are observed to litter more in a littered environment
than in a clean environment
\citep{cialdini1990focus}. Similarly, when empirical and normative
expectations are in conflict, people tend to follow the empirical
expectations. One potential explanation for this is that people are
rarely punished when everyone is engaging in the same behavior
\citep{bicchieri2009right}. Little is known about what happens when
personal norms are in conflict with 
descriptive or injunctive
norms, or when personal
normative beliefs are in contrast with empirical and normative
expectations. The example that we mentioned in Section
\ref{se:moral_pref} of a vegan who does not eat food containing
animal-derived ingredients while realizing that it is an injunctive
norm to do so suggests 
that, at least in some cases, personal judgments about what is right
and what is wrong represent the dominant motivation for behavior. 

In the context of the games considered in this review, some 
experimental work points towards a significant role of personal norms, 
at least in one-shot and anonymous interactions. For example,
\cite{capraro2018right} created a laboratory 
setting in which the personal norm was pitted against the descriptive
norm in the trade-off game, 
and found that participants tended to follow the personal norm. More recently, \cite{catola2021personal}
showed that personal norms predict cooperative behavior in the
public-goods game better than social norms do. The role of personal norms 
was also highlighted by \cite{bavsic2021personal}. They found that
both personal and social norms shape behavior in the dictator and
ultimatum games, which 
led them to propose a utility function that takes into account both
types of norms,  
as reviewed in Section \ref{se:moral_pref}.
In
any case, we believe that an important direction for future empirical
%
research is an
exploration of how people resolve norm conflicts. This may
be a key step in allowing us to create a new generation of utility
functions that take into account the relative effect of different
types of norms.

Another major direction for future work is the
exploration of how the heterogeneity in individual personal norms
affects economic choices.
People differ
in their judgments about what they consider right and wrong.
For example, some people 
embrace utilitarian ethics, which dictates that the action selected
should maximize total welfare and minimize total harm
\citep{mill2016utilitarianism,bentham1996collected}. Others
embrace \emph{deontological ethics}, according to which the rightness or
wrongness of an action is entirely determined by whether the action
respects certain moral norms and duties, regardless of its
consequences \citep{kant2002groundwork}. It has been
suggested that people's personal norms can be 
decomposed into fundamental dimensions, although
there is some debate about the number and the characterization of these dimensions. 
According to moral foundations theory, there are six dimensions:
care/harm, fairness/cheating, loyalty/betrayal,
authority/subversion, sanctity/degradation, and liberty/oppression
\citep{haidt2004intuitive,graham2013moral,iyer2012understanding,haidt2012righteous}%
; according to morality-as-cooperation theory, there are seven dimensions:
helping kin, helping your group, reciprocating, being brave, deferring to superiors, 
dividing disputed resources, and respecting prior possession
 \citep{curry2016morality,curry2019good,curry2019mapping}. 
Each individual assigns different weights to these dimensions.
These weights have been shown to play a key
role in determining a range of important characteristics, including
political orientation \citep{graham2009liberals}. 
There has been very little work exploring the link between 
moral dimensions and prosocial behavior. Some preliminary evidence 
does suggest that different forms of prosocial behavior may be associated
with different moral foundations, not necessarily correlated between themselves. 
For example, while both dictator game giving and
ultimatum game rejections appear to be associated with moral preferences, giving 
appears to be primarily driven by the fairness dimension of morality \citep{schier2016moral},
whereas ultimatum game rejections seem to be primarily driven by the ingroup
dimension \citep{capraro2021moral}. It is worth noting that the fairness and the ingroup dimensions 
are not correlated between themselves \citep{haidt2004intuitive,graham2013moral,iyer2012understanding,haidt2012righteous}.
Therefore, these preliminary results speak in favor of the multidimensionality
of prosociality and may provide a rationalization for the result by \cite{chapman2018econographics} that 
the giving cluster of social preferences is not correlated with the punishment cluster.
As said, this line of research has just started.\footnote{There is some work exploring the link between moral 
foundations and cooperation in the prisoner's dilemma \citep{clark2017behavioral}. Moreover, a working paper by \cite{bonneau2021different} explores the role of different
moral foundations in explaining altruistic behavior in different forms of the 
dictator game, including the dictator game with a take option and
the dictator game with an exit option.}
Understanding 
the link between moral dimensions and prosocial behaviors
 is a necessary step for building
models that can explain human behavior with greater
precision. 

Since our moral preferences are clearly affected by our social
interactions, and it is well known that the structure of an
individual's social 
network affects preferences and outcomes in general
\citep{easley2010networks}, we believe that another important line of
research is how moral preferences are shaped by social connections.
For example, it is known that cooperation is strongly affected by the structure of the social network.
Hubs in such networks can act as strong cooperative centers and exert a positive influence on 
the peripheral nodes \citep{santos_pnas06}. The ability to break ties
with unfair or exploitative partners and make new one's with those of
better reputation also favorably affects cooperation
\citep{perc_bs10}. More recent research has shown that other forms of
moral behavior, such as truth-telling and honesty,
are also strongly affected by the properties of
social networks \citep{capraro_pre20}. And a case has been made for
further explorations 
being much needed along these lines \citep{capraro_fp18}, for example, by
studying how network structure affects different types of moral
behavior, including equity, efficiency, and trustworthiness. 

Another
line of research involves language-based preferences. To the
extent that people do have language-based preferences, it would
be useful to be able to predict 
how people will behave in an economic decision problem described
by a language. A relatively new area of research in computational
linguistics may be relevant in this regard.   \emph{Sentiment 
analysis} (e.g.,
\citet{pang2002thumbs,pang2004sentimental,esuli2007sentiwordnet}) 
aims to determine the attitude of a speaker or a writer to
        a topic from the information contained in a document. For
        example, we may want to determine 
 the feelings of a reviewer about a movie from his
review. 
Among other things, sentiment analysis attempts to associate to a
description (in a given context) its \emph{polarity}, that is, 
a number in the interval $[-1,1]$
expressing a positive, negative, or indifferent sentiment.
One 
could
perhaps use sentiment analysis to define a
utility function by taking the polarity of the description of
    strategies into account.
    The idea is that people are
    reluctant to perform actions that evoke negative sentiments, like
        stealing, but are eager to perform
 actions that evoke positive sentiments; the utility function could
 take this into account (in addition to taking into account the monetary
 consequences of actions).
  Being able to 
associate
utility to words would also allow us to 
measure the explanatory power
of language-based models. Experimental Regularity 7 shows that language-based models explain 
behavior in some games better than models based solely on monetary 
outcomes. However, to the best of our knowledge, there has been no econometric study
measuring the exploratory power of language-based models. 
We have only scratched the surface here of potential research
directions. We believe that economics is taking brave steps into uncharted territory here. New ideas will be needed, and bridges to other fields will need to be built. The prospects are most exciting!

\bibliographystyle{aea}
\bibliography{moral_preferences}

\end{document}